\documentclass[oldversion]{aa}
\usepackage{times}
\usepackage{amssymb}
\usepackage{mathptmx}
\usepackage{natbib}
\bibpunct{(}{)}{;}{a}{}{,}  
\usepackage[]{graphicx}
\usepackage{url}
\usepackage{color}
\usepackage{comment}
\usepackage{txfonts}
\usepackage{txfonts}

\begin{document}

\title{Jet-driven AGN feedback on molecular gas and low star-formation efficiency in a massive local spiral galaxy with bright X-ray halo \thanks{Based on data obtained with ALMA through program 2019.1.01492.S and with IRAM through program 080$-$19.}}
\author{N.~P.~H.~Nesvadba\thanks{email:
nicole.nesvadba@oca.eu}\inst{1}, A.~Y.~Wagner\inst{2}, D.~Mukherjee\inst{3}, A.~Mandal\inst{3}, R.~M.~J.~Janssen\inst{4,5}, H.~Zovaro\inst{6}, N.~Neumayer\inst{7}, J.~Bagchi\inst{8}, G.~Bicknell\inst{6}}
\institute{
  Universit\'e de la C\^ote d'Azur, Observatoire de la C\^ote d'Azur,
  CNRS, Laboratoire Lagrange, Bd de l'Observatoire, CS 34229, 06304
  Nice cedex 4, France
\and 
University of Tsukuba, Center for Computational Sciences, Tennodai 1-1-1, 305-0006, Tsukuba, Ibaraki, Japan
\and
Inter-University Centre for Astronomy and Astrophysics, Post Bag 4, Pune - 411007, India
\and
Jet Propulsion Laboratory, California Institute of Technology, 4800 Oak Grove Dr., Pasadena, CA 91109, United States
\and
Department of Astronomy, California Institute of Technology, 1216 E California Blvd., Pasadena, CA 91125, United States
\and 
Research School of Astronomy and Astrophysics, The Australian National University, Canberra, ACT 2611, Australia
\and
Max-Planck-Institut für Astronomie, K\"onigstuhl 17, 69117, Heidelberg, Germany
\and
Department of Physics \& Electronics, CHRIST (Deemed to be University), Hosur Road, Bengaluru 560029, India}
\authorrunning{Nesvadba et al.}
\titlerunning{Molecular disk in J2345-0449}

\date{Received  / Accepted }

\abstract{It has long been suspected that powerful radio sources may lower the efficiency with which stars form from the molecular gas in their host galaxy, but so far, alternative mechanisms, in particular related to the stellar mass distribution in the massive bulges of their host galaxies, are not ruled out. We present new,
  arcsecond-resolution ALMA CO(1--0) interferometry, which probes the spatially resolved, cold molecular gas in the nearby ($z=0.0755$), massive ($M_{stellar}=4\times 10^{11} M_{\odot}$), isolated, late-type spiral galaxy 2MASSX~J23453269$-$044925, which is outstanding for having two pairs of powerful, giant radio jets, and a bright X-ray halo of hot circumgalactic gas. The molecular gas is in a massive ($M_{gas}=2.0\times 10^{10} M_{\odot}$), 24~kpc wide, rapidly rotating ring, which is associated with the inner stellar disk. Broad ($FWHM=70-180$ km s$^{-1}$) emission lines with complex profiles associated with the radio source are seen over large regions in the ring, indicating gas velocities that are high enough to keep the otherwise marginally Toomre-stable gas from fragmenting into gravitationally bound, star-forming clouds. About 1-2\% of the jet kinetic energy are required to power these motions. Resolved star-formation rate surface densities derived from GALEX fall factors 50$-$75 short of expectations from the standard Kennicutt-Schmidt law of star-forming galaxies, and near gas-rich early-type galaxies with signatures of star formation lowered by jet feedback. We argue that radio AGN feedback is the only plausible mechanism to explain the low star-formation rates in this galaxy. Previous authors have already noted that the X-ray halo of J2345$-$0449 implies a baryon fraction that is close to the cosmic average, which is very high for a galaxy. We contrast this finding with other, equally massive, and equally baryon-rich spiral galaxies without prominent radio sources. Most of the baryons in these galaxies are in stars, not in the halos. We also discuss the implications of our results for our general understanding of AGN feedback in massive galaxies.}

\keywords{Galaxies -- ... -- ...}

\maketitle
\section{Introduction} 
\label{sec:introduction}

The energy injection from active galactic nuclei (AGN) into the interstellar gas of their host galaxy is now a widely accepted mechanism of galaxy evolution, supported mainly by the observation of fast outflows in galaxies associated with powerful AGN \citep[e.g.,][]{morganti05,nesvadba06,nesvadba08,hardcastle12,mahony13,nyland13,harrison14,woo16,karouzos16,fiore17,nesvadba17a,mukherjee18,komossa18,menci19,zovaro19,santoro20}. About 30\% of nearby massive radio galaxies, which host considerable amounts of warm or cold molecular gas \citep[][]{ocana10, ogle10, dicken14}, have also been found to have less intense star formation for their molecular gas content than other types of more typical star-forming galaxies \citep[e.g.,][]{nesvadba10, ogle10,alatalo11, lanz16}. \citet{sabater19} found that nuclear radio activity seems to be near-ubiquitous in massive, early-type galaxies with stellar masses, $M_{stellar} > 10^{11}$ M$_{\odot}$, making feedback from radio
sources in AGN host galaxies a very interesting mechanism to explain the dearth of star formation in many massive galaxies at low and intermediate redshifts.

In spite of the current popularity of these models, AGN feedback is not the only mechanism that has been proposed in the literature to explain the generally low star-formation rates in gas-rich early-type galaxies. Some authors point out, that it would be more natural to explain the low star-formation rates in parts of the galaxy population through the structural properties of the host galaxies or underlying dark-matter halos. Proposed mechanisms include the long gas cooling times of gas heated by the virial shock as it falls into massive dark-matter halos \citep[e.g.,][ but see also, e.g., \citealt{keres05}]{white78, birnboim07, ogle19}, the high angular momentum of gas accreted onto massive rotating galaxies \citep[e.g.,][]{renzini20,peng20}, and the high stellar mass surface densities and spheroidal shape of early-type galaxies and massive bulges \citep[][]{dekel14, tacchella16, martig09, martig13}. Low star-formation rates may also be a  consequence of strong bars in spiral galaxies \citep[][]{fraser-mckelvie18}. Moreover, massive galaxies reside preferentially in dense environments, where galaxies may lose their gas content rapidly through ram-pressure stripping and interactions with neighboring galaxies, perhaps even before being accreted onto a massive central galaxy \citep[][]{donnari20}. In particular such mechanisms would have the advantage that they are universal byproducts of the mass assembly of massive galaxies themselves. Feedback, in contrast, is an additional process which must be assumed to be at work. It is also limited to the activity periods of the supermassive black hole, which can be very short compared to a Hubble time \citep[e.g.,][]{eilers17}. These scenarios do pose an important challenge to AGN feedback models, because they imply that low star-formation rates and AGN activity in galaxies, even when driving outflows, may be coincidental without requiring a causal, physical link.  

A simple, straight forward way to break the degeneracy between radio AGN feedback and those other mechanisms would be to study the interstellar medium and star formation in a massive, gas-rich radio galaxy, which shows signatures of feedback, but which has neither bars, nor a rich environment, nor a bulge or unusually high stellar mass-surface densities. Such galaxies are however very hard to come by, as powerful radio-loud AGN activity is almost inextricably related to massive bulges. For example, \citet{tadhunter16} find that $<5$\% of radio galaxies in the 2~Jy and 3~CR radio samples are hosted by spiral galaxies (typically spirals with massive bulges) over their entire range in stellar mass. Their radio sources are usually compact and populate the low-power end of the radio luminosity function. The fraction of spiral galaxies decreases further when focusing on radio sources with more complex morphologies. For example, \citet{singh15} identified only four galaxies with two pairs of radio jets in a sample of 187 000 spiral galaxies, for which they identified radio counterparts in the NVSS and FIRST radio surveys. Such giant, double-lobed radio sources are generally considered evidence of sustained, repeated, AGN activity over long timescales.

Rare, outstanding galaxies can be extremely valuable to understand a more general astrophysical mechanism, if their properties allow us to isolate an individual process that cannot be studied very easily in the dominant galaxy population because of the co-existence of other, potentially rivalling mechanisms producing a similar effect. Showing that radio jets can lower star formation rates in even a single galaxy where competing mechanisms can be ruled out, would demonstrate that jets are in principle capable of doing so also in the overall galaxy population. 

Here we present an analysis of newly obtained, arcsecond-resolution ALMA interferometry of CO(1--0) of the nearby, z=0.0755, massive spiral galaxy 2MASSX~J23453269$-$044925 (J2345$-$0449 hereafter) with a stellar mass of $M_{stellar}=4\times 10^{11}$ M$_{\odot}$ and a radio power of $P_{1.4\ GHz}=2.5\times 10^{24}$~W~Hz$^{-1}$ at 1.4~GHz \citep[][]{bagchi14,walker15}. The galaxy was first mentioned in the literature by \citet{machalski07}, and then further discussed by \citet{bagchi14}, who obtained low-frequency radio observations with the GMRT and optical longslit spectroscopy along the major and minor axis of the 35\arcsec\ large stellar disk. They identified two pairs of bright, giant radio jets extending over 390~kpc and 1.6~Mpc, respectively, suggesting that radio AGN activity has been on-going for long timescales.  

J2345-0449 is particularly suited to investigate the impact of powerful radio activity onto the star formation in its host galaxy, because it shows no indication of several other, potentially rivaling quenching mechanisms: It has a small ($<$1.5~kpc) pseudo-bulge contributing at most 15\% of the total stellar mass, which probably formed from secular processes \citep[][]{bagchi14, kormendy04}. It also resides in a poor environment, with only one intermediate-mass neighbor at a distance of 240~kpc \citep[][]{bagchi14}. The nuclear spectrum is at the same time that of a Low-Excitation Radio Galaxy (LERG), a Weak-Line Radio Galaxy \citep[WERG][]{tadhunter98}, and a LINER \citep[][]{bagchi14}, ruling out the presence of strong radiative feedback from a bolometrically bright AGN. Star formation is too weak to drive an outflow \citep[][]{walker15}. 

\citet{walker15} showed with Chandra X-ray imaging that the galaxy is surrounded by a bright X-ray halo, potentially with cavities along the jet axis, which they attribute to hot gas heated by the radio source. They estimated a jet kinetic energy of $2\times 10^{44}$ erg s$^{-1}$, enough to produce a bright, hot gaseous halo as observed with Chandra out to 80~kpc. The total mass of this halo was recently estimated with XMM-Newton spectroscopy \citep{mirakhor20}, and is $M_{gas,\ X}=8.25^{+1.62}_{-1.77}\times 10^{11}$ M$_{\odot}$ out to the virial radius, corresponding to a baryon fraction of J2345$-$0449 that is close to the cosmic average, implying that the underlying dark-matter halo would have retained nearly all its baryons. This is unusual compared to other spiral galaxies with stellar masses of few $10^{11}$ M$_{\odot}$, which are rather poor in hot circumgalactic gas \citep{li18}, but rich in colder baryons: \citet{posti19} find that the most massive galaxies in the SPARCS survey \citep[][]{lelli15} may have stellar-to-dark-matter mass ratios
that are near the cosmic average, placing them amongst the sites of most efficient star formation known. 

\citet{dabhade20b} recently observed J2345$-$0449 as part of a sample of galaxies with giant radio jets with the IRAM 30-m telescope, and found that J2345$-$0449 hosts about $2\times 10^{10}$ M$_{\odot}$ of cold molecular gas, which is highly unusual for giant radio galaxies, which are generally gas-poor. This result was obtained independently from us and published well after our own IRAM single-dish, and ALMA interferometric datasets had already been taken.

How did J2345$-$0449 obtain such unusual properties? \citet{borzyszkowski17} and \citet{jackson20} have proposed, based on cosmological models of galaxy evolution, that locations near, but outside of massive galaxy clusters may favor the occurrence of massive spiral galaxies with particularly quiescent evolutionary histories. They reason that the node in the Cosmic Web traced by the cluster would serve as an attractor to neighbouring galaxies which would otherwise have produced merging events destroying the large stellar disk. J2345$-$0449 only has three neighbors at $<500$~kpc distance, but lies in the wider vicinity of the massive, X-ray galaxy cluster RBS~2042 at 2.8~Mpc distance. This may perhaps explain why this galaxy could retain its late-type spiral structure until today. The reason why the supermassive black hole in this galaxy has developed such an outstanding radio source is currently unclear. \citet[][]{bagchi14} discuss several hypotheses. This does not hinder us to investigate the effect of the radio source on the interstellar medium and the on-going star-formation based on the galaxy properties that we can observe in-situ.

We will in the following describe our observations of the molecular
gas and stellar continuum of J2345$-$0449, and discuss our results in the light of AGN feedback and star formation in massive galaxies.  We will start with presenting the new data set and ancillary, publicly available data that we used for our analysis in Section~\ref{sec:observations} before deriving additional properties of the stellar population that will be needed for the subsequent analysis (Section~\ref{sec:massivespiral}). In Section~\ref{sec:ring} we will discuss the properties of the newly discovered molecular ring, the gas kinematics and gas mass surface density, before showing in  Section~\ref{sec:agnkinematics} how they relate to the kpc-scale structure of the radio source seen in FIRST. In Section~\ref{sec:sflaw} we demonstrate that the resolved star-formation rate densities and gas mass surface densities place J2345$-$0449 firmly below the sequence formed by spiral galaxies with normal star formation in the Kennicutt-Schmidt and alaternative star-formation laws, and into the realm of gas-rich early-type radio galaxies with signatures of heating of molecular gas through radio jets. In the same section we also examine the rotational support of the gas with help of the Toomre criterion. In Section~\ref{sec:discussion} we discuss the implications of our results and previous X-ray observations for AGN feedback from radio jets and alternative mechanisms. We summarize our work in Section~\ref{sec:summary}.

Throughout the paper we use the flat $\Lambda$CDM cosmology from \citet{planck18} with $H_0=67.4$ km s$^{-1}$ Mpc$^{-1}$, $\Omega_M=0.315$, and $\Omega_{\Lambda}=1-\Omega_M$. At $z=0.0755$ the luminosity distance is $D_L= 354.6$~Mpc, $1.49$~kpc are projected onto one arcsecond.

\begin{figure}
  \centering
  \includegraphics[width=0.49\textwidth]{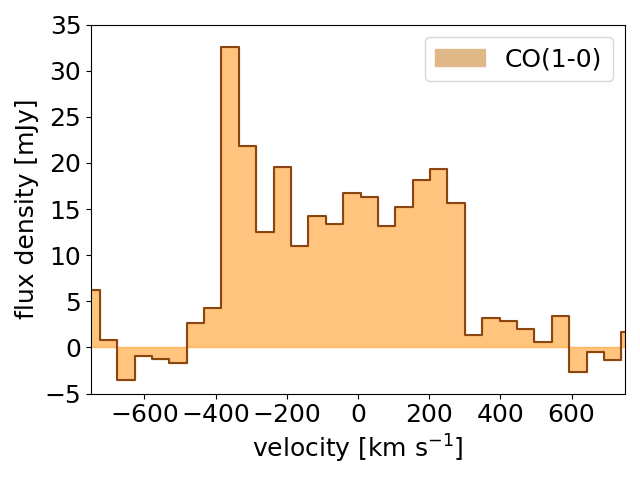}
  \caption{\label{fig:emir} Our CO(1--0) single-dish spectrum of
    J2345-0449 obtained with the EMIR wide-band receiver at the 30-m
    telescope of IRAM in program $080-19$.}
\end{figure}

\section{Observations and data reduction}
\label{sec:observations}

\subsection{IRAM/EMIR single-dish observations}
We initially obtained 3~mm single-dish data of J2345-0449 with the
wide-band millimeter receiver EMIR at the 30-m telescope of IRAM in
one short, 7.5~hrs run on 3~July, 2019 under program-ID 080-19 (PI Nesvadba). The receiver was tuned to 107.179~GHz, the expected frequency of CO(1--0) for a redshift
z=0.0755. We used the FTS and WILMA backends with Wobbler switching
throws of 60\arcsec, which is significantly larger than the diameter
of our source. To point the telescope we used blind offsets from a
nearby quasar, and performed a pointing about every 2~hrs. The
telescope was refocused after 4~hrs.

Data were calibrated at the telescope and reduced with the CLASS
package of the GILDAS software of IRAM \citep[][]{gildas13}. Each
individual scan was inspected by eye, and we used simple first-order
polynomials to remove the baselines, after carefully masking the
spectral range where we expected CO(1--0) to fall. We used the values
given on the EMIR
website\footnote{http://www.iram.es/IRAMES/mainWiki/Iram30mEfficiencies}
to translate the measured brightness temperatures into units of flux
density (Jy). Fig.~\ref{fig:emir} shows that the line is well detected
at the expected frequency, with a clear, although asymmetric,
double-horn profile of FWHM=684$\pm$18 km s$^{-1}$ and an integrated
line flux of $12.6\pm 0.8$~Jy km s$^{-1}$. A similar spectrum has
recently been obtained by \citet{dabhade20b} who measured a somewhat
greater integrated line flux of 14.0 Jy km s$^{-1}$. The difference
might be due to calibration uncertainties. We will in the following
use our own measurement as a single-dish reference when discussing the
new ALMA interferometry.

\subsection{ALMA CO(1--0) interferometry}
Our ALMA data were taken with the 12~m-array on 17~October, 2019,
through observing program 2019.1.01492.S (PI Nesvadba). We used 47~antennae in the
3~mm band at a representative frequency of 107.179~GHz with the 43-3
configuration. The on-source observing
time was 1179.36~sec, which we obtained under average conditions with
a column of precipitable water vapor, $pwv=1.75$~mm. The quasar
J0006$-$0623 was used as bandpass and flux calibrator. We used the Common Astronomy Software Application (CASA) version 5.6 to reduce the data, applying the automatic flagging of visibilites, to calibrate the bandpass, phase, amplitude and flux. Synthesized beam-deconvolved data cubes were constructed with CLEAN, using briggs weighting with $robust=0.5$. The spatial sampling of the final data cube is 0.2\arcsec, and spectral channel sizes are 11 km s$^{-1}$. The rms is ~332~$\mu$Jy~bm$^{-1}$. The beam size of the synthesized data cube is 1.32\arcsec$\times$1.07\arcsec, with a position angle of 273.2$^\circ$. The detailed analysis of this data set is described in Section~\ref{sec:ring}.

\subsection{Ancillary data}
We complement these new proprietary data with several publicly
available data sets of J2345$-$0449. To study the optical and
near-infrared stellar continuum in our source, we obtained u, g, r, i,
and z-band imaging from the SDSS data release 16, as well as near-infared J, H, and Ks band imaging from 2MASS. Moreover, we retrieved the fully reduced and calibrated GALEX Near and Far-UV imaging and the WISE infrared imaging from the respective archives. 

J2345$-$0449 also has 1.4~GHz imaging available that was obtained at
the Very Large Array as part of the FIRST survey in the B-array
\citep[][]{becker95}. The beam size is 5\arcsec, and the depth at the position
of J2345$-$0449 is rms=0.37~mJy bm$^{-1}$. The core and lobes of
J2345$-$0449 are well resolved. The lobes span 1.6~Mpc, making
J2345$-$0449 one one of the largest radio sources on the sky
\citep[][]{bagchi14}. The core has an integrated flux
density of 4.4~mJy in FIRST and extends from the nucleus to about
10\arcsec\ north north west (Fig.~\ref{fig:fwhmjet}).

\section{J2345$-$0449 as a massive spiral galaxy}
\label{sec:massivespiral}

\subsection{Bulge-disk decomposition}
\label{ssec:sigmastars}

We will in the following give separate mass estimates for the bulge
and disk of J2435$-$0449. \citet{bagchi14} already
decomposed the optical continuum of J2345$-$0449 into a bulge and a
disk component, finding bulge sizes of $r_e=0.9-1.25$~kpc, i.e.,
0.6\arcsec-0.8\arcsec, and a low contribution of the bulge to the total
optical luminosity of 14-18\%, depending on the band.

However, they did not quantify the stellar mass in each
component. This is best done in the near-infrared, where the
mass-to-light ratio of stellar populations does not depend as strongly
on the star-formation history as in the optical
\citep[e.g.,][]{bell03}. We therefore provide a new estimate, using
the J-band image from 2MASS. J-band is favored over the K-band because of the higher image quality.

To separate the bulge from the disk component, we construct an azimuthally averaged surface-brightness profile with
the Python implementation of the Ellipse package
\citep[][]{jedrzewski87}, and use the LM-fitting routine of
\citet{markwardt12} to model this profile with two Sersic functions,
each convolved with a two-dimensional Gaussian to take into account
the seeing. 
\begin{figure*}
\centering
  \includegraphics[width=0.49\textwidth]{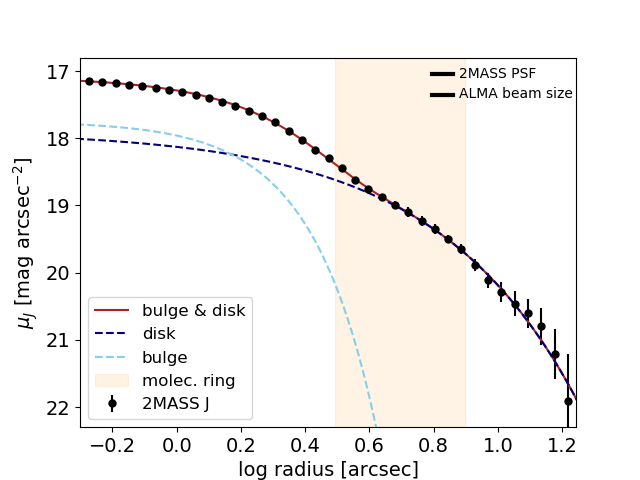}
  \includegraphics[width=0.49\textwidth]{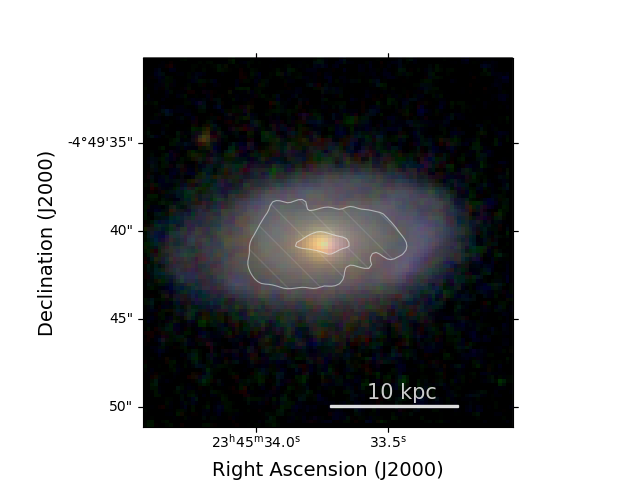}
  \caption{
  \label{fig:sbprofile} {\it (left:)} Azimuthally averaged
    surface brightness profile of J2345$-$0449 extracted from the
    2MASS J-band image. The dark and light blue components show individual fits to the disk and bulge component, respectively. The red line shows the combined fit. Both Sersic components were also convolved with a Gaussian kernel with $FWHM=$0.96\arcsec, corresponding to the size of the seeing disk in the 2MASS data. The light orange area shows the radii where CO(1--0) line emission is detected in the ALMA data. {\it (right:)} SDSS three-color image of
    J2345$-$0449, showing the galaxy in the g, r, and i-bands. The
    hatched area shows where CO(1--0) line emission is detected with
    ALMA at $\ge 3\times$ the rms. North is up, east to the left.
}
\end{figure*}

The result is shown in Fig.~\ref{fig:sbprofile}. The
surface-brightness profile of J2345$-$0449 in the J-band is well
fitted with two Sersic components, that are convolved with a Gaussian kernel of $FWHM=0.96$, to approximate the size of the seeing disk in the 2MASS image. For the central bulge component, we find a Sersic parameter, $n=0.47$, and for the disk component, $n=1.0$. Corresponding effective radii are 1.7\arcsec\ and 8.1\arcsec, respectively. These Sersic parameters are within the range of those previously found by \citet{bagchi14} in the optical, however, the effective radii are systematically larger. The seeing is comparable in the optical and near-infrared data, however, the spatial sampling in the 2MASS data is much coarser, with 1\arcsec\ pixel size, and the signal-to-noise ratio is lower, in particular in the outer disk. This does not affect the main purpose of our fit, which is to constrain the observed radius where the bulge blends into the disk, but it might affect the detailed fit results, in particular the measured radii. In the following we will consider all light emitted from within 3\arcsec\ of the nucleus as coming from the bulge, and light from outside that radius as coming from the disk.    

\subsection{Stellar masses, and stellar mass surface densities}
\label{ssec:stellarmass}

Massive spiral galaxies have likely formed most of their stars in rapid bursts of star formation at high redshift \citep[][]{zhou20}. 
To translate the J-band magnitudes and surface
brightnesses of J2345$-$0449 into estimates of stellar mass and mass surface density,
we therefore used the population synthesis models of \citet{bruzual03} to
estimate the mass-to-light ratios, $\Upsilon$, of a fiducial 10-Gyr old stellar population formed in a fairly short (200~Myr) period of constant star formation of 50~M$_{\odot}$ yr$^{-1}$, as is frequently observed in
massive galaxies at redshifts z$\ga$2. For these parameters we find   $\Upsilon=0.5~M_{\odot}/L_{\odot}$.
\begin{figure*}
  \centering
  \includegraphics[width=0.49\textwidth]{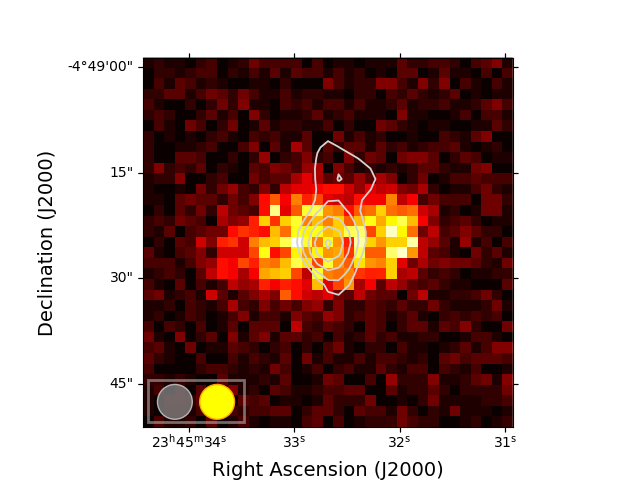}
  \includegraphics[width=0.49\textwidth]{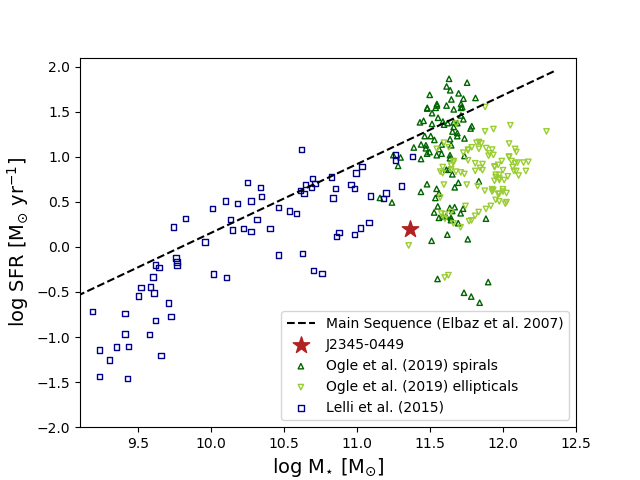}
  \caption{\label{fig:galex} 
  GALEX near-UV image of
    J2345-0449, showing extended, diffuse emission associated with the
    stellar disk. No nuclear point source can be discerned. The contours
    show the radio source observed with FIRST. The FWHM size of the beam
    of FIRST, and the point-spread function of GALEX, are show in
    the lower left corner. Both are 5\arcsec.}
\end{figure*}

At a luminosity distance of $D_L=354.6$~Mpc, J2345$-$0449 has an
absolute J-band magnitude of 12.7~mag, corresponding to a stellar mass
of M$_{\star,J}=3.6\times 10^{11}$ M$_{\odot}$. This is somewhat less
than the previous estimate of M$_{\star,K}=4.6\times 10^{11}$ M$_{\odot}$ found by \citet{walker15}, who used the integrated K-band
magnitude of J2345$-$0449 and a fiducial mass-to-light ratio of
$\Upsilon_K = 0.78~M_{\odot}/L_{\odot}$ taken from \citet{bell03}. A
systematic discrepancy of 25\% is acceptable for the purpose of our
analysis, and corresponds, e.g., to the difference in the J-band mass-to-light ratio of a stellar population with 10 and 12~Gyrs of age, respectively. Providing a more robust estimate would require detailed
population synthesis modeling of the star-formation history, as will
be possible when our VLT/MUSE observations will become available, which
have been delayed due to the Covid-19 shutdown of the VLT.

Individual integrated magnitudes of the bulge and disk component of
J2345$-$0449 are 13.9~mag and 13.1~mag, respectively, corresponding to
stellar masses of $M_{bulge}=1.2\times 10^{11}$ M$_{\odot}$ and
$M_{disk}=2.5\times 10^{11}$ M$_{\odot}$, respectively, assuming that
both have old stellar populations. Corresponding stellar mass surface
densities are $\Sigma_{stellar}= 2\times 10^3$~M$_{\odot}$~pc$^{-2}$ and
$\Sigma_{stellar}= 2\times 10^2$~M$_{\odot}$~pc$^{-2}$ for the bulge and
disk, respectively.

Values of $\Sigma_{bulge}= {\rm few} \times 10^3~M_{\odot}$~pc$^{-2}$
are typical for bulges of spiral galaxies, including of lower mass,
and for early-type galaxies \citep[e.g.,][]{kauffmann03,fang13,gonzalezdelgado15}. Likewise, stellar mass
surface densities of $\Sigma_{disk}= {\rm few}\times
100~M_{\odot}$~pc$^{-2}$ are not untypical for late-type galaxies over
large ranges in stellar mass \citep[e.g.][]{kauffmann03,fang13,gonzalezdelgado15}. Thus, J2345$-$0449 may
have an extraordinarily large integrated stellar mass for being a spiral
galaxy, however, the mass surface densities are not
outstanding. J2345$-$0449 is more massive than other spiral galaxies,
because it is more extended, not because it has a higher stellar mass
surface density in either the bulge or the disk.

\subsection{Star formation}
\label{ssec:starformation}

Estimating star-formation rates in AGN host galaxies is notoriously
difficult, as most tracers can be contaminated by the AGN. We
integrated the far-UV flux in the GALEX image of J2345$-$0449, finding
$F_{FUV}=92.1\pm1.1\ \mu$Jy, which corresponds to an integrated
star-formation rate of $SFR=1.25\pm0.1~M_{\odot}$ yr$^{-1}$. We derived this estimate with  the relationship of \citet{kennicutt98}, but adopted the today more widely used Kroupa \citep[][]{kroupa01} stellar initially mass function instead of that of Salpeter \citep[][]{salpeter59}, as \citet{kennicutt98} did initially. The uncertainty includes only the error of
the measurement, not the systematic uncertainties. \citet{dabhade20b}
estimated $SFR=2.95$~M$_{\odot}$ yr$^{-1}$ from the 22~$\mu$m flux
measured with WISE, whereas \citet{walker15} found SFR=1.6~M$_{\odot}$
yr$^{-1}$ from the same GALEX UV continuum that we also used. Both  adopted a Salpeter initial mass function. We estimated a star-formation rate by combining the UV and IR measurements into a single estimator, using the relationship of \citet{leroy08}. They approximate the star-formation rate by setting  
$SFR= 8.1\times 10^{-2}\ I_{FUV}\ +\ 3.2^{+1.2}_{-0.7}\times 10^{-3}\ I_{24}$. We use the FUV intensity measured with GALEX, $I_{FUV}$, and the 22~$\mu$m intensity corresponding to the magnitude given by \citet{dabhade20b}, $m_{22\mu m}=7.135$~mag. This gives a very similar value to our above estimate using GALEX alone, $SFR = 0.9-1.4\  M_{\odot}~$~yr$^{-1}$, with a best estimate of $SFR=1.1$~M$_{\odot}$ yr$^{-1}$. This suggest that the UV does not miss significant amounts of star formation in this galaxy. 

\begin{figure*}
  \centering
  \includegraphics[width=0.33\textwidth]{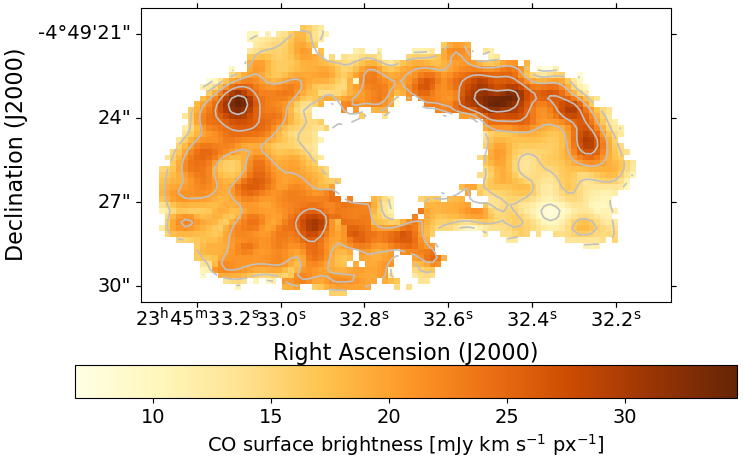}
  \includegraphics[width=0.33\textwidth]{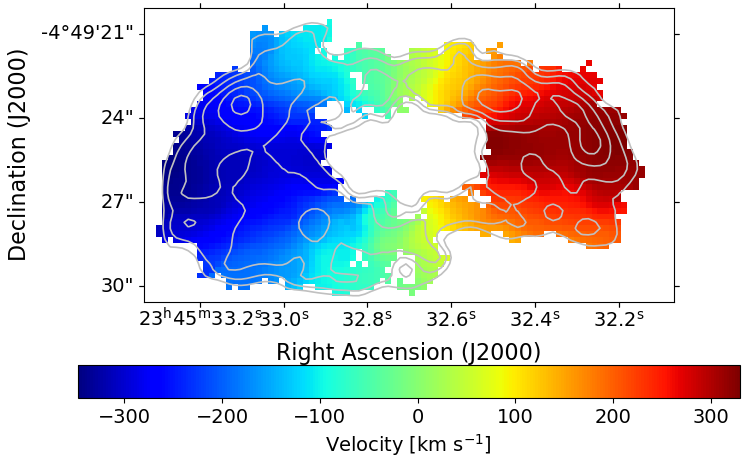}
  \includegraphics[width=0.33\textwidth]{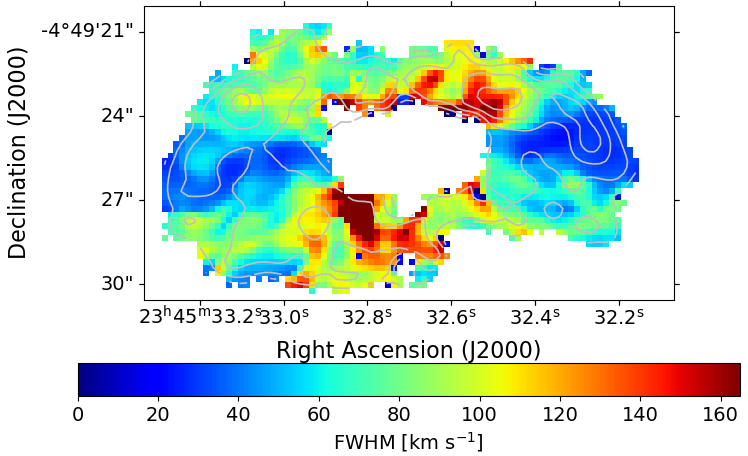}\\
  \caption{\label{fig:almamaps} 
  {\it left to right:} CO(1--0) emission-line morphology, and maps of relative velocity and FWHM
    of CO(1-0) in J2345$-$0449. Velocities are given relative to
    z=0.0755.}
\end{figure*}

The GALEX and WISE imaging also allow us to investigate where the star
formation in J2345$-$0449 is located relative to the molecular gas,
radio jet, and stellar disk. The left panel of Fig.~\ref{fig:galex}
shows the near-UV image of J2345$-$0449, which is well resolved at
PSF$=$5\arcsec, and has a relatively uniform surface-brightness
distribution without signature of a nuclear point source. Diffuse UV
emission is detected over an area with a deconvolved size of
28.6\arcsec~$\times$~14.0\arcsec, corresponding to
43~kpc~$\times$~21~kpc, i.e., over most of the size of the 50~kpc large stellar disk. Peak surface brightnesses are reached in the
eastern and western regions of the disk, and do not coincide with the
position of the radio source shown as contours. From the deconvolved
size estimate of the disk and the integrated SFR=1.25~M$_{\odot}$
yr$^{-1}$, we estimate an average star-formation rate surface density
of $\Sigma_{SFR}=1.9\times 10^{-3}$~M$_{\odot}$ yr$^{-1}$
kpc$^{-2}$. Local minima and maxima correspond to
$SFR=1.3-5.6\times10^{-3}$ M$_{\odot}$ yr$^{-1}$ kpc$^{-2}$. 

The spatial resolution of the WISE~22~$\mu$m image (not shown), which has a point spread function of $FWHM=11.9$\arcsec, is even less than that of GALEX, making it difficult to use to estimate star-formation rate surface densities. However, WISE does resolve the disk of the galaxy, with a measured major-axis size of $FWHM=28.3$\arcsec, corresponding to 25.7\arcsec\ after correcting for the PSF. This is comparable to the spatial extent of the UV emission, suggesting that both FIR and UV are emitted from the same regions of the galaxy. 

In the right panel of Fig.~\ref{fig:galex} we show where J2345$-$0449
falls relative to the main sequence of star-forming galaxies of
\citet{elbaz07}, and other samples of spiral and elliptical galaxies
taken from \citet{ogle19} and \citet{lelli15}, which both include very spiral galaxies with $M_{stellar} \ge 10^{11}$ M$_{\odot}$. J2345$-$0449
falls 0.95~dex below the ridge-line of \citet{elbaz07}, and near the
low end of star-formation activity of massive spiral galaxies
generally. Interestingly, as also shown in Fig.~\ref{fig:galex}, the
star-formation rate is also at the low end of the range spanned by
elliptical galaxies in the same mass range. With SFR=1.25~M$_{\odot}$
and M$_{stellar}=3.6\times 10^{11}$ M$_{\odot}$, the specific
star-formation rate of J2345$-$0449 is sSFR=$3.3\times 10^{-3}$
Gyr$^{-1}$.

\subsubsection{Radio source}
\label{ssec:radio}

\citet{singh15} found only four spirals with multiple pairs of radio jets in a sample of over 187 000 spiral galaxies. Current samples of very massive (M$_{stellar}>10^{11}$ M$_{\odot}$) spiral galaxies are very small, and do not allow for a systematic analysis, however, we investigated the radio
morphologies of all massive spiral galaxies in the samples of
\citet{lelli15} and \citet{ogle19} in the NVSS and FIRST surveys 
(32 sources in total), and did not identify a single galaxy with a radio
morphology akin to that of J2345$-$0449. Galaxies with radio AGN are
usually point sources in FIRST at 5\arcsec\ beam size. Sources which
are extended have morphologies that follow the stellar disk, sometimes
even showing the same spiral structure. Their radio luminosities are also consistent with being related to star formation, not AGN
activity. It is difficult to draw firm quantitative conclusions from
these sources, since they are not taken from a complete
sample. However, they are at least consistent with powerful radio jets
being rare in spiral galaxies with stellar masses $>10^{11}$
M$_{\odot}$, further highlighting the exceptional nature of J2345$-$0449 already pointed out by \citet{singh15}. 

\section{The molecular ring of J2345$-$0449}
\label{sec:ring}
\subsection{CO(1--0) morphology}

Fig.~\ref{fig:almamaps} shows the maps of CO(1--0) that we obtained
with ALMA. The molecular gas is in a large ring with an outer radius
of 7.9\arcsec$\times$4.4\arcsec\ (11.8~kpc$\times$6.6~kpc), and has a
strong gradient of projected velocities of 676$\pm$25~km~s$^{-1}$
increasing from West to East. For an inclination angle of $59^\circ$
\citep[][see also \S\ref{ssec:emkin}]{bagchi14}, this corresponds to a
deprojected, intrinsic velocity offset of 789$\pm$25 km s$^{-1}$. We saw in Section~\ref{ssec:stellarmass} that the size of the bulge in the J-band is somewhat larger, 3\arcsec, but can be accomodated within the measurement uncertainties set by the large pixel scale and size of the seeing disk (both are 1\arcsec). 

We do not detect any CO(1--0) line emission inside a region of
2.8\arcsec$\times$1.5\arcsec\ from the nucleus, corresponding to a
diameter of 4.2~kpc$\times$2.2~kpc at the redshift of J2345$-$0449,
and down to a flux limit of r.m.s.=332~$\mu$Jy~bm$^{-1}$. This is more
than twice the effective radius of the stellar bulge in the SDSS
r-band image, $r_e=1.25$~kpc, the largest size measured by
\citet{bagchi14}. We can therefore safely conclude that the molecular
gas is associated with the inner disk of J2345$-$0449, and avoids the
bulge. This is a significant difference to quasars, for which disks of molecular gas are found to have much smaller radii of $\le~2$~kpc, e.g.,
\citet{molina21}.
The morphology is fairly irregular, showing several bright clouds with
sizes of $\le$1\arcsec (1.5~kpc), and potentially, gas that is
following the spiral arm structure in the north-western part of the
ring. The radial width of the ring varies from a minimum of
2.0\arcsec\ (3.0~kpc) along around $PA=200^\circ$ in the southern part
to 5.7\arcsec\ (8.5~kpc) along around $PA=300^\circ$ (position angle
PA is measured counterclockwise from north towards east). Given the
irregular overall morphology, this cannot be attributed only to
projection effects. The ring extends to a smaller radius than that of
the stellar disk seen with the SDSS (which has a radius of
16.9\arcsec\ in the $i$-band image of the SDSS down to 3$\sigma$,
Fig.~\ref{fig:sbprofile}). This corresponds to 25.2~kpc at the redshift of J2345$-$0449.

The total line flux is $I_{CO,ALMA}=14.9\pm 0.5$~Jy km s$^{-1}$,
comparable to what was previously found in the single-dish
observations of \citet[][$I_{CO}=14.0$~Jy~km~s$^{-1}$]{dabhade20b} and
our own measurements ($I_{CO}=12.6\pm 0.8$ Jy km s$^{-1}$). We suspect that
the reason for the difference is the large diameter of the ring, which
we trace out to 19.6\arcsec, comparable to the half-power beamsize of
the IRAM 30-m telescope of 22.9\arcsec\ at the observed
frequency. Line emission at large radii would be attenuated by the
lower efficiency in the outer regions of the beam of the 30~m
telescope of IRAM.

To test more quantitatively whether this is a viable explanation, we
convolve the ALMA map with a two-dimensional Gaussian distribution
with FWHM widths corresponding to the beam size of the 30-m telescope,
and centered on the position at which we pointed the telescope. We
find a total flux of $I_{CO}=13.0$~Jy~km~s$^{-1}$. This is less than
$1\sigma$ from the integrated flux of $I_{CO}=12.6\pm0.8$~mJy we
measure with the 30~m telescope, and slightly lower than the 14 Jy km
s$^{-1}$ found by \citet{dabhade20b}, and suggests that the large size
of the ring compared to the telescope beam is indeed the likely cause
of the lower flux measurement in the single-dish data. For our ALMA
observations this implies that we have recovered at least 94\% of the
single-dish flux, if the \citet{dabhade20b} estimate is closer to the
actual flux of J2345-0449. If our estimate is more accurate, it would
suggest a systematic uncertainty in the relative flux calibration of
about 3\%, which is also very plausible. In either case, we can
conclude that the amount of missing flux is small, and we will
neglect it in our following analysis.
\begin{figure*}
  \centering
  \includegraphics[width=0.32\textwidth]{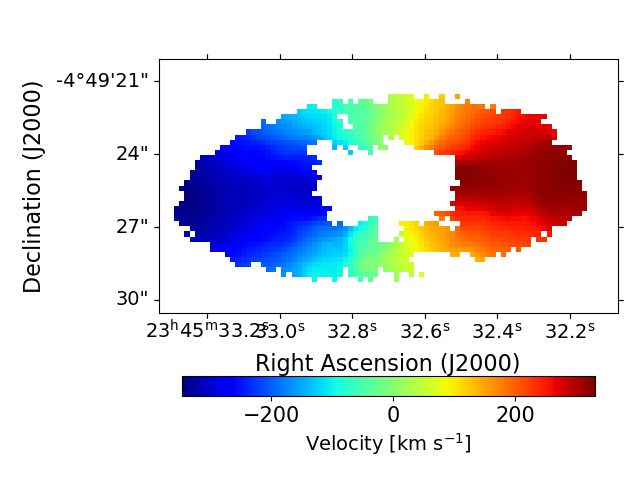}
  \includegraphics[width=0.32\textwidth]{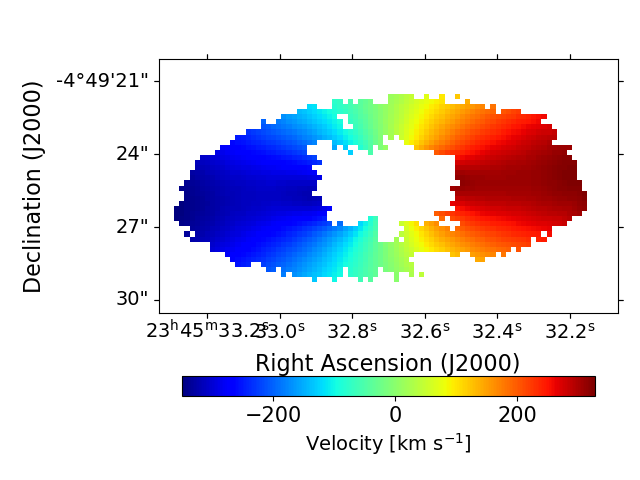}
  \includegraphics[width=0.32\textwidth]{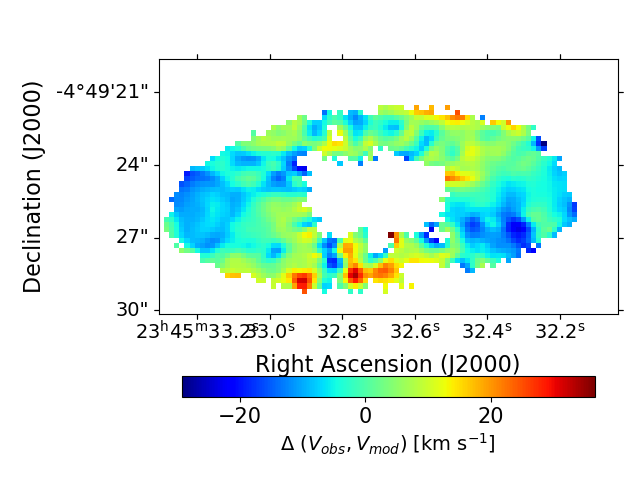}
\caption{
\label{fig:kinemetry} Results of the
  Fourier analysis of the velocity distribution in the ring, using
  Kinemetry \citep{krajnovic06}. {\it left:} Observed velocity
  map. {\it center:} Output of the best-fit model. {\it right:}
  Residuals in the measured velocity map after subtracting the model
  shown in the center.}
\end{figure*}

\subsection{Gas kinematics}
\label{ssec:emkin}

To extract spatially-resolved information about the molecular gas kinematics in this galaxy we constructed spectra from square apertures with  1.0\arcsec$\times$1.0\arcsec\ in size in a data cube with 11~km s$^{-1}$ wide spectral channels, comparable to the beam
size, and as broad as possible to measure line widths for the molecular gas. We then fitted each spectrum with a single Gaussian component. This was done for all pixels with signal-to-noise ratio $SNR>3$ in an emission-line map constructed from directly integrating
all flux in the redshift range where line emission was detected. Typical signal-to-noise ratios per spectral channel at the line peak were between 3 and 6 for the broad, and 8 to 11 for the narrow emission lines. 

Fig.~\ref{fig:almamaps} shows the results: In the left panel, the
integrated line flux derived from the single-Gaussian fits can be
seen, indicating several massive gas clouds embedded into an extended,
irregular ring. The central panel shows the local gas velocities as
measured from the centroids of the single-component Gaussian line
profiles, given relative to a redshift $z=0.0755$
\citep[][]{bagchi14}. The right panel shows the FWHM line widths in
each aperture.

\subsection{Fourier-analysis of the large-scale velocity field}
\label{ssec:kinemetry}
The gas velocities in J2345$-$0449 are very regular and increase smoothly from east to west as
expected from rotation within the gravitational potential of the host
galaxy (central panel of Fig.~\ref{fig:almamaps}).
We used Kinemetry \citep[][]{krajnovic06} to describe the velocity
distribution more quantitatively. Kinemetry performs a fifth-order Fourier analysis of the kinetic moments of the
line-of-sight velocity distribution of a gas disk or ring along
concentric ellipses, thereby representing a generalization of the
classical surface photometry. It is a standard tool to analyze data
cubes obtained with integral-field spectrographs. The ring of
J2345$-$0449 is well spatially resolved with a total of about 30
individual spatial resolution elements, so that a detailed analysis as
done with Kinemetry can be expected to yield meaningful results.

Kinemetry requires that the centroid position and inclination angle of
the disk are fitted externally. To do so, we used the gas morphology,
which we fitted with the Python implementation of the IRAF task {\sc
  ellipse} \citep[][]{jedrzewski87}. We find that the ring is centered
on position (J2000) RA=23:45:32.708, Dec=-04:49:25.42, where ALMA also
shows a faint ($S_{110GHz} = 1.3\pm0.34$~mJy) radio continuum point source, presumably the
nucleus. The position angle on the sky is $PA=94^\circ$, and the
inclination angle $59^\circ \pm 2^\circ$, the same values previously
also found for the stellar component by \citet{bagchi14}. This is
another indication that most of the gas in the disk, as measured from
the line centroids, rotates largely unperturbed within the
gravitational potential of the host galaxy.

A comparison of the observed and modeled velocity distributions in the
ring (corresponding to the velocity map) is shown in
Fig.~\ref{fig:kinemetry}. Differences are small and difficult to
discern by eye. The right panel of the figure shows the residual map
obtained from subtracting the model from the measured velocity
distribution.  Residuals are less than about $\pm10$ km s$^{-1}$ in
most of the disk, or $1.5$\% of the measured total velocity offset,
with somewhat greater offsets ($\ge 30$ km s$^{-1}$) along the outer
rim. Parts of these differences may also come from complex line
profiles which may affect the centroid position of the single-Gaussian
fits. 
We have also compared offsets between higher kinetic
moments, finding that model and observations agree generally within
about 2\%. To summarize, gas velocities in the ring as probed by the
line centroids do not show significant deviations from rotational
motion within the gravitational potential of J2345$-$0449.

\subsection{Position-velocity diagram}
\label{ssec:pvdiag}
\begin{figure}
  \centering
  \includegraphics[width=0.49\textwidth]{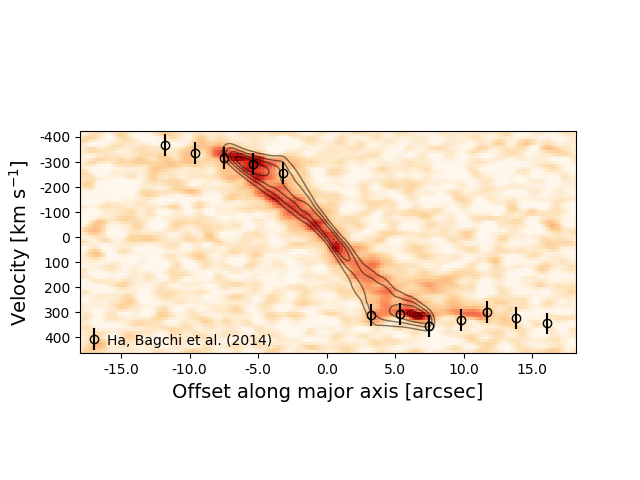}
  \caption{\label{fig:pvdiag} CO(1--0) position-velocity diagram of
    the ring in J2345-0449 along the morphological and kinematic major
    axis of the galaxy at $PA=94^\circ$. Contours show the best-fit
    disk model from {\sc kinemetry} \citep[][]{krajnovic06}. The
    slight offset on the Eastern branch comes from the irregular
    surface-brightness distribution of the ring. Black dots with error
    bars show the velocities derived by \citet{bagchi14} from longslit
    spectroscopy of H$\alpha$ along the major axis of the stellar
    disk.}
\end{figure}

Another way to visualize the good agreement of the gas kinematics in
the ring of J2345-0449 with rotational motion is by analyzing the
position-velocity (PV) diagram shown in
Fig.~\ref{fig:pvdiag}. Distances are measured from the nucleus of the
galaxy along the morphological and kinematic major axis of the
galaxy at $PA=94^\circ$. The Figure shows a near-linear decline in
observed velocity from east to west out to $R=5$~kpc on either side,
where velocities flatten, suggesting that the ring samples the
rotation curve at radii just outside the turnover radius. Contours show the
PV-diagram of the best-fit model obtained with Kinemetry for
comparison (\S\ref{ssec:kinemetry}). The slight mismatch between the
observed and modeled ring in the eastern hemisphere is caused by the
irregular surface-brightness distribution of the ring, which was not
taken into account in the modeled data set. Overall, the figure shows
few residuals in the data with the exception of faint line emission
to the outermost radii in the western part of the disk.

We further overplot the rotation curve obtained
previously by \citet{bagchi14} from H$\alpha$ longslit spectroscopy
extracted along the morphological major axis of the disk. Both gas
components show very similar kinematics, suggesting that both arise
from the same structure. However, H$\alpha$ samples the gas out to
larger radii. The warm ionized gas seems also to extend somewhat
further inward that the molecular gas. However, \citet{bagchi14}
report a dearth of H$\alpha$ within a radius of 1.5\arcsec\ from the
nucleus, where they detect only [NII]$\lambda\lambda$6548, 6583. It is
therefore possible that most of the warm ionized gas is also in a ring
rather than a disk. The faint line emission found in the extreme outer
western part of the ring follows the velocities measured in H$\alpha$,
and may be a faint molecular component emitted from the same
environment.

The circular velocity of the ring in J2345$-$0449 obtained from this
fit is $369\pm25$ km s$^{-1}$, derived from the projected (observed)
velocity of $\Delta v / 2$ = 316~km s$^{-1}$ at the turnover radius of
5.2~kpc, and corrected for an inclination of $i=59^\circ$. The
turnover radius is probably an upper limit, given that it falls near
the inner edge of the ring.

Fig.~\ref{fig:pvdiag} also shows that this velocity is consistent
with the velocity previously found by \citet{bagchi14} from H$\alpha$
at the same radius. Their asymptotic circular velocity estimate,
$V_c=371\pm26$ km s$^{-1}$ is however larger, since H$\alpha$ samples
the velocity curve out to larger radii, and velocities continue to
increase slowly, in particular on the eastern side of the ring.
The enclosed dynamical mass within the turnover radius, $M_{dyn}= V^2\ R / G$, is $M_{dyn}=1.6\times10^{11}$ M$_{\odot}$, corresponding to about 44\% of the integrated stellar mass $3.6\times 10^{11} M_{\odot}$ (\S\ref{ssec:stellarmass}). This is not much larger than the stellar mass previously estimated for the bulge (\S\ref{ssec:stellarmass}), showing that the dark-matter content in the central regions of J2345$-$0449 is low. 

\subsection{Line widths}
In spite of the smooth velocity map, the CO line profiles are very
irregular, and full-widths at half maximum (FHWMs) of the CO emission
lines change rapidly throughout the ring, increasing from a minimum of
about 20 km s$^{-1}$ in the eastern and western parts of the ring to
more than 120 km s$^{-1}$ in the northern and southern region as seen
in projection (Fig.~\ref{fig:almamaps}). Overall, we observe
significant line broadening over large ranges in position angle, from $PA=-70^\circ$ to $PA=75^\circ$, and from $PA=113^\circ$ to $PA=246^\circ$. Generally,
line widths are greater in the inner regions of the ring
(Fig.~\ref{fig:almamaps}).

\begin{figure*}
  \centering
  \includegraphics[width=0.32\textwidth]{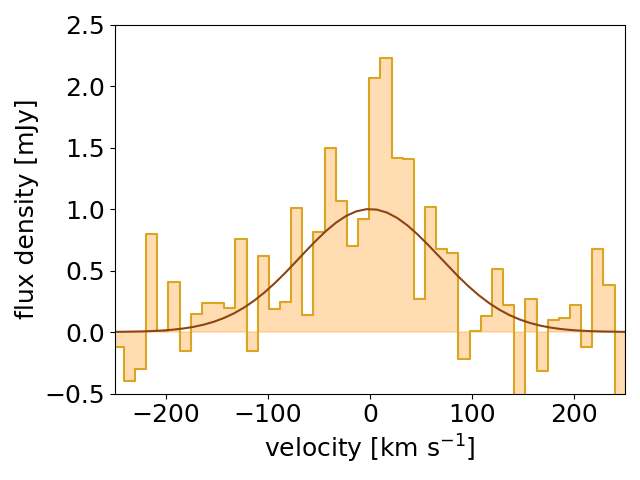}
  \includegraphics[width=0.32\textwidth]{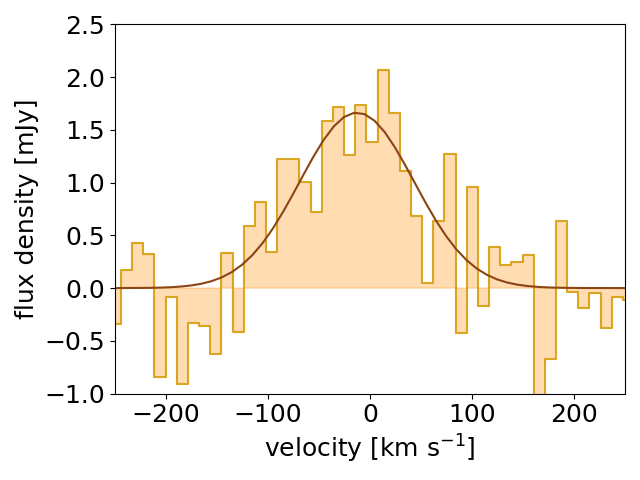}
  \includegraphics[width=0.32\textwidth]{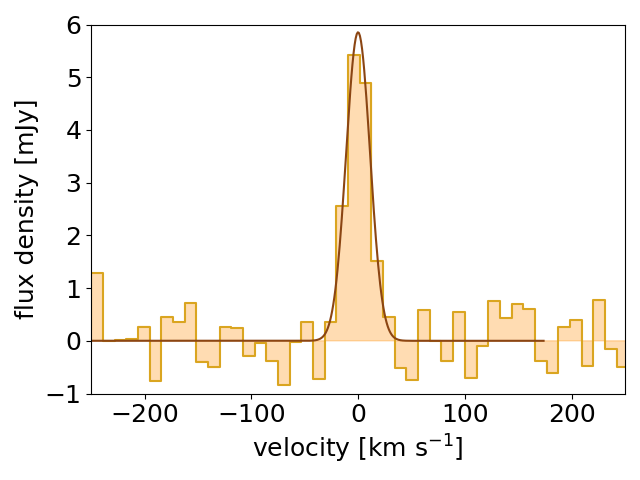}
\caption{
\label{fig:spectra} Local molecular gas kinematics extracted in 1\arcsec\ apertures from the ALMA CO(1--0) data cube inside (left and center) and outside (right) the interaction zone with the radio jet.}
\end{figure*}

A closer look at the line profiles in individual pixels also shows
that the gas kinematics are strongly perturbed. We identify secondary
components in regions of narrow line widths, and in the transition
regions between narrow and broad line emission. In the regions of
broadest widths, however, the line profiles appear fairly smooth,
consisting of a single component, at least at the spatial and spectral
resolution of our data (Fig.~\ref{fig:spectra}).

\subsection{Integrated molecular gas mass and gas mass surface densities}
\label{ssec:sigmagas}
We use Equation~3 of \citet{solomon97} to estimate a molecular gas
mass from the CO flux measurement, by setting $L^\prime = 3.25\times
10^7\ I_{CO}\ (\nu/(1+z))^{-2}\ D_L^2\ (1+z)^{-3}\ {\rm K}\ {\rm km}\ {\rm
  s}^{-1}\ {\rm pc}^2$, where $\nu$ is the observed frequency, $z$ the
redshift, and $D_L$ the luminosity distance of our target. For an
observed integrated CO(1--0) line flux of $I_{CO}=14.9\pm 0.5$ Jy km
s$^{-1}$, we find L$^\prime$ = $4.7\times 10^9$ K km s$^{-1}$
pc$^2$. For a CO-to-$H_2$ conversion factor of $\alpha_{CO}=4.3\ {\rm
  M}_{\odot}$ [K km s$^{-1}$ pc$^2$]$^{-1}$, commonly used for
nearby spiral and early-type galaxies \citep[e.g.,][]{bolatto13}, we find a total
molecular gas mass of M$_{H2}=2.0\times 10^{10}$ M$_{\odot}$. We did not include a correction for He and other metals, if we had, this mass estimate would be 36\% higher, M$_{H2}=2.7\times 10^{10}$ M$_{\odot}$.

This is at par with cold molecular gas masses previously measured from
CO(1-0) in the most gas-rich, nearby, classical radio galaxies like
3C~293, 3C~84, or 4C~12.50. However, these are typically S0 galaxies
rather than classical elliptical galaxies. The most gas-rich
elliptical radio galaxies have cold molecular gas masses of at most
few $10^9$ M$_{\odot}$ \citep[][see also the compilation of
  \citealt{lanz16}, and references therein]{ocana10}, factors of a few
greater than typical gas masses in quiescent radio galaxies
\citep[][]{tadhunter16}. In exceptional cases, gas-rich
elliptical galaxies can have warm molecular gas masses from shocked
gas, presumably by the AGN, equal to or even above $1\times 10^{10}$
M$_{\odot}$ in the most extreme cases
\citep[][]{ogle10}. A similar mass range, $M_{gas,cold}=1\times 10^8 -
4\times10^{10}$ M$_{\odot}$, is also suggested by Herschel/SPIRE
measurements of the dust mass in radio AGN host galaxies drawn from
the 2~Jy sample by \citet{dicken14}, assuming gas-to-dust ratios of
140.

Since we have spatially well resolved measurements of the molecular
gas in the ring, we can also estimate the local gas mass surface
densities in individual 1.5~kpc apertures, corresponding to the
1\arcsec\ beam size. Assuming that a single CO-to-H$_2$ conversion
factor of $\alpha_{CO}=4.3\times$ M$_{\odot}$ [K km s$^{-1}$
  pc$^2$]$^{-1}$ applies throughout J2345$-$0449, the average of the
molecular gas mass surface densities for the 381~kpc$^2$ surface of
the ring is $\Sigma_{gas,ave}=55$~M$_{\odot}$ pc$^{2}$, with a median
of $\Sigma_{gas}=83$ M$_{\odot}$ pc$^{-2}$, and a maximum surface
density of $\Sigma_{gas}=196$~M$_{\odot}$ pc$^{-2}$ reached in the
brightest clouds.

These gas mass surface densities are at the upper end, but still
comparable to those found in other spiral galaxies,
and at the lower end of starbursts \citep[e.g.,][]{kennicutt98}. They
are also in the range of those of massive early-type radio galaxies
with large amounts of molecular gas studied by \citet[][]{ogle10} and
\citet[][]{lanz16}.

\section{Signatures of AGN-driven gas kinematics}
\label{sec:agnkinematics}

Fig.~\ref{fig:almamaps} showed that the molecular gas in J2345$-$0449
is in a large molecular ring associated with the stellar disk of the
host galaxy, which is rapidly rotating, and shows strongly increased
line widths up to $FWHM\approx180$ km s$^{-1}$ in the northern and southern part of the ring. We will now discuss this line broadening in more
detail, show that it is outstanding compared to other spiral galaxies
in the same mass range, and demonstrate that the radio jet is the most
likely cause.

\subsection{Radio jets as origin of the broad line emission}

\begin{figure}
  \centering
  \includegraphics[width=0.49\textwidth]{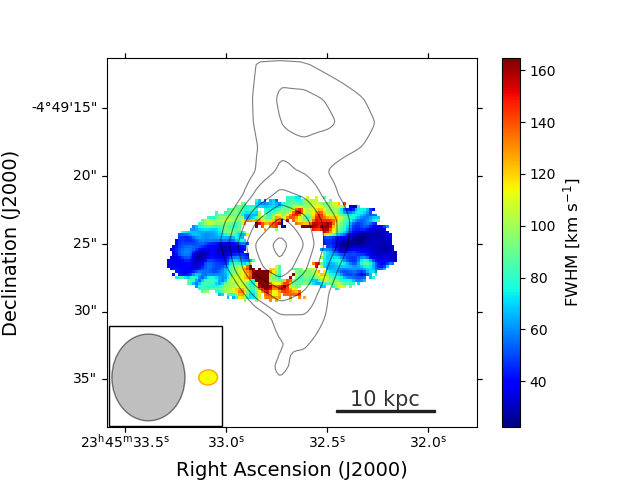}
  \caption{
  \label{fig:fwhmjet} Map of FWHMs 
    obtained with ALMA shown as color image, with the 1.4~GHz radio
    morphology observed with FIRST overplotted as contours. The region
    of strong line broadening of CO coincides well with the position
    angle of the radio source.}
\end{figure}

In Fig.~\ref{fig:fwhmjet} we show the map of FWHMs of the CO(1-0) line
emission with the small-scale morphology of the radio jet overplotted
as contours. The radio image was taken from the VLA FIRST survey
\citep[][]{becker95} obtained at 1.4~GHz with 5\arcsec\ beam size. In spite of the relatively large beam size compared to
the ALMA data, we can clearly see that the 
radio source is extended along an axis roughly going from
north to south. The region over which the broad CO line emission is
detected clearly coincides with the region of the ring intercepted by
the radio source. This makes it very plausible that the radio emission
from the AGN is also the main responsible for the broadening of the
emission lines.

Moreover, the observed CO(1--0) line widths of FWHM$=70-180$ km
s$^{-1}$ in this region are outstanding compared to other massive
spiral galaxies without prominent radio jets. There are not many such
galaxies in the literature with stellar masses $\ge 10^{11}$
M$_{\odot}$ and detailed analyses of the line widths measured from
spatially resolved CO data sets. Existing data include NGC~5055,
NGC~7331, NGC~2841, and NGC~3521 from the THINGS survey \citep[][]{tamburro09},
which have circular velocities between 200 and 300 km s$^{-1}$ and
velocity dispersions that are similar to those measured in HI, $\sigma\approx 10$ km s$^{-1}$
corresponding to $FWHM=25-35$ km s$^{-1}$.  \citet{tamburro09} find a
maximum of $\sigma \le 60$ km s$^{-1}$ in in NGC~5055 within the
central regions of the galaxy, with smaller values further outside.

We caution that the data of these galaxies were obtained
at $\ge$10\arcsec\ beam size, much larger than the beam size with which we
observed J2345$-$0449. Thus, these dispersions are probably much
more affected by beam smearing effects. They should therefore be
treated as conservative upper limits. Nonetheless, these observations
allow us to conclude that maximal intrinsic velocity dispersions,
$\sigma$, in J2345$-$0449 caused by gravitational motion or other
processes in common with massive disk galaxies without prominent radio
source, are probably below $25-30$ km s$^{-1}$, corresponding to FWHMs
which are below 70 km s$^{-1}$. Line widths $>$70 km s$^{-1}$ must
therefore be caused by another process, and the spatial association
with the radio source points clearly to the most likely culprit, even
more so as we do not see any evidence for other processes that could
potentially increase the line widths in J2345$-$0445, like a strong
bar or interactions with a massive neighbor. Moreover,
\citet{walker15} have already argued why star formation in
J2345$-$0449 is not strong enough to drive an outflow.

We have also used a toy model to investigate how much of this line
broadening can be attributed to blending of gas motions within the
rapidly rotating ring in our own data. We used the modeled velocity
field to construct an artificial data cube with the same
signal-to-noise ratio and beam size as observed in our data, and
extracted the emission-line kinematics in precisely the same was as
previously done for the data. We found that beam smearing doubles the
widths of lines with intrinsic $FWHM=10$ km s$^{-1}$, whereas lines
with an intrinsic width of $FWHM= 25$ km s$^{-1}$ are broadened by at
most $2$ km s$^{-1}$. This is negligible compared to the observed
range in line widths of up to 180 km s$^{-1}$ and shows that most of
the line widths observed in the northern and southern hemisphere of
the ring cannot be explained by beam smearing.

In total, $L^\prime_{CO10,~broad}= 11.2$~K~km~s$^{-1}$~pc$^2$ of the
line flux in J2345-0449 are emitted from gas with $FWHM>70$~km
s$^{-1}$, and $L^\prime_{CO10,~narrow}=3.7$~K~km~s$^{-1}$~pc$^2$ from
gas that is more quiescent. Assuming that similar CO-to-H$_2$
conversion factors apply to both regions, molecular gas masses are
$M_{gas,~broad}=1.6\times 10^{10}$~M$_{\odot}$ and
$M_{gas,~narrow}=0.5\times 10^{10}$~M$_{\odot}$,
respectively. Corresponding gas mass surface densities are
$\Sigma_{gas,~broad}=85$~M$_{\odot}$~pc$^{-2}$ and
$\Sigma_{gas,~narrow}=75$~ M$_{\odot}$~pc$^{-2}$, respectively, for the
perturbed and the quiescent gas. It is of course possible, that the
perturbed mass would be lower because another CO-to-H$_2$ conversion
factor applies in the perturbed gas
\citep[e.g.][]{alatalo11}. Adopting instead $\alpha_{CO}=\ 0.34\ 
M_{\odot}$~[K~km~s$^{-1}$~pc$^2$]$^{-1}$ as a lower limit, appropriate for
optically thin gas at low excitation temperature \citep[][]{bolatto13},
we would find that $M_{gas,broad,opt~thin}=1.2\times 10^9$ M$_{\odot}$ of
molecular gas are perturbed by the radio source. This would still be a
significant fraction of the total molecular gas content of this
galaxy, and comparable to the total molecular gas mass in many
radio-loud early-type galaxies \citep[e.g.,][]{ogle10}.

\subsection{Search for outflow signatures} 
\label{ssec:outflows}
To investigate the properties of coherent non-circular gas motions over scales of several kpc, as could be expected from filamentary gas at the edge of an expanding bubble, an outflow, or galactic fountains, 
we have tried to perform a multi-component fit to the line profiles,
and also adopted non-parametric methods. We have also performed fits with priors, e.g.,
on the modeled large-scale velocity maps from Kinemetry \citep[][see
  also \S\ref{ssec:kinemetry}]{krajnovic06}, trying to isolate a
systemic gas component from additional lines with non-rotational
kinematics like outflows. 

However, although the line profiles
extracted from larger apertures clearly indicate that such components are present, likely indicating that the gas is in multiple molecular clouds or filaments with distinct kinematic properties, the signal-to-noise ratios in individual small apertures are not sufficient to probe the global kinematic properties of multiple components robustly and in more detail. Tracing such components over larger areas consisting of multiple spectral elements is also hindered by the blending of rotational and non-rotational components, whose velocity offset is often commensurate with the steep velocity gradient from rotational motion over few pixels. Probing non-rotational components would therefore require deeper observations with higher spatial and spectral resolution than the 20~min of on-source observing time we have obtained for this program, and which was only meant to yield a detection of CO line emission in this galaxy.

We have in particular searched for evidence of outflows that would be fast enough to escape either from the molecular ring and underlying stellar disk, or even the galaxy altogether, in either case depleting the molecular reservoir for on-going star formation. While the complex line profiles make it difficult to identify individual outflow components, pronounced line wings extending to velocities greater than the circular velocity of the disk should be more easily seen. We did not identify any such component down to a peak rms = 1.2~mJy in our data cube. For an assumed line width of FWHM=180~km~s$^{-1}$, the largest seen in this galaxy, this would correspond to a 3$\sigma$ upper limit of 0.69~Jy~km~s$^{-1}$, corresponding to 5\% of the integrated CO(1-0) flux of the galaxy, or $9.7\times 10^8$ M$_{\odot}$ in case that the same CO-to-H$_2$ conversion factor applies that we adopted for the disk. This has the character of an upper limit, as molecular outflows have previously been found to be optically thin \citep[][]{alatalo11, dasyra16}, which would lower the potential outflowing mass by about an order of magnitude. 

We have very little constraints on the motion of other gas phases in this region of J2345$-$0449. The only constraints we are aware of are from \citet{bagchi14}, who show the flat radial velocity profile of H$\alpha$ extracted from a 1.5\arcsec\ slit aligned with the morphological minor axis of the galaxy, indicating no significant offsets from the systemic redshift. It is unclear whether this implies the absence of line broadening, or perhaps a wind of ionized gas, however, since they make no statement about the line widths and
profiles. A more complete picture will emerge once that we will have MUSE data of J2345$-$0449 at hand, which were delayed due to the Covid-19 shutdown.

\subsection{Kinetic energy}
\label{ssec:kinenergy}
We can give a rough estimate of the kinetic energy corresponding to
the line broadening in this region, by setting $E_{kin} =
3/2\ M\ v^2$, i.e., assuming isotropic gas broadening in three
dimensions. Summing the CO line flux over all regions where $FWHM>70$
km s$^{-1}$ suggests that a total of $1.6\times 10^{10}$ M$_{\odot}$
is being stirred up by interactions with the radio source,
corresponding to about 75\% of the total molecular gas mass in the
ring (Section~\ref{ssec:sigmagas}). Using the local Gaussian line widths measured in each spatial
element to estimate the kinetic energy, we find a total of
$E_{kin}=1.3\times 10^{57}$ erg in the gas.

Can the required energy to power these gas flows be produced by the radio
source in J2345$-$0449? \citet{walker15} used the \citet{cavagnolo10}
relationship between observed jet power at 1.4~GHz and the kinetic
energy traced through X-ray cavities in galaxy clusters to estimate
the kinetic power of the radio jet in J2345$-$0449,
$P_{kin,jet}$. They found $P_{kin,jet} = 2\times 10^{44}$ erg
s$^{-1}$, corresponding to a total energy ejection of $7\times
10^{58}$ erg for a fiducial jet lifetime of 10~Myr. They consider such
a lifetime, which they derive from the 780~kpc size of each of the outer jets
and an assumed maximal jet advance speed of $0.25\ c$ as a plausible
lower limit. Detailed analyses of other giant radio galaxies find that
jet advance speeds can be as low as $0.01\ c$ \citep[e.g.,][]{jamrozy05}, increasing this age
estimate to as much as 250~Myr, in which case the total kinetic energy carried by the radio source would increase to $1.7\times 10^{60}$ erg. We also note that the \citeauthor{cavagnolo10} estimate was derived from the work done by FRI radio galaxies in galaxy clusters to inflate X-ray cavities. While this estimate is frequently used to estimate jet kinetic energies, it may systematically underestimate the true jet kinetic energy, which is however very difficult to constrain directly \citep[e.g.,][]{mukherjee18b}. Nonetheless, this represents a very large energy output, comparable to that of the AGN in massive central cluster galaxies, in spite of being emitted from a black hole residing in the center of a late-type spiral galaxy. 

Over 10~Myrs of jet lifetime, 2\% of the jet kinetic energy would need to be deposited in the gas to explain the observed line widths. This is comparable to the values found by \citet{mukherjee16} from relativistic hydrodynamic simulations \citep[see also][]{mukherjee17}. For comparison, if we had adopted the common velocity estimate of $v_t
= \Delta v + 2\sigma$ initially proposed by \citet{rupke05}, a
commonly adopted measure of outflow velocity in quasars, the kinetic
energy estimate would be $5.3\times 10^{57}$ erg, corresponding to 8\%
of the jet kinetic energy. We caution, however, that this estimate is
only valid under certain assumptions, which are likely not met in this
case. It provides an interesting quantity, however, to compare with
more generic sets of AGN probed with spatially unresolved data.

\subsection{Alternative scenarios}

We have just argued that the radio jet itself carries more kinetic
energy than required to explain the gas kinematics in J2345$-$0449, in
agreement with the results of hydrodynamic simulations. However, to
demonstrate that this is a unique interpretation, it is also necessary
to rule out alternative scenarios. 

The kinetic power of the jet is much higher than that carried by other mechanisms that
could cause feedback in J2345$-$0449. Direct constraints on the
presence of the bolometric luminosity of the AGN are scarce in the
literature, however, \citet{bagchi14} do show the spectrum of the
bulge, which clearly falls into the LERG \citep[][]{buttiglione10} or
WLRG \citep[][]{tadhunter98} regime of radio galaxies characteristic
for sources with very weak (if any) bolometric AGN
emission. [OIII]$\lambda\lambda$4949,5007 line emission is not seen in
this spectrum, and we can use the empirical relationship of
\citet{heckman04}, to estimate the bolometric AGN luminosity by
setting $L_{bol,AGN}=3500\times L_{[OIII]5007}$. With the upper limit
of $L\le1\times 10^{40}$ erg s$^{-1}$ given by \citet{bagchi14} for
optical line emission, this would correspond to
$L_{bol,AGN}=3.5\times 10^{43}$ erg s$^{-1}$. Radiative processes
like, e.g., radiation pressure, would therefore be at least about a
factor~10 less energetic than the kinetic power carried by the radio
source.

\citet{walker15} already argued that star formation would also be too
weak to cause significant feedback in J2345$-$0449. They estimate a
supernova rate of 0.12 M$_{\odot}$ yr$^{-1}$. Adopting that each
supernova ejects $10^{49}$ erg s$^{-1}$ in kinetic power
\citep[e.g.,][]{dallavecchia08}, this would correspond to a kinetic energy
injection of $4\times 10^{40}$ erg s$^{-1}$ from star formation, more
than three orders of magnitude less than from the radio source. We
conclude, in agreement with \citet{walker15}, that the radio source is
by far the dominant source of kinetic energy in J2345$-$0449.

\section{Star-formation law}
\label{sec:sflaw}

\subsection{J2345$-$0449 as an outlier in the standard Kennicutt-Schmidt law and alternative relationships}

Numerous studies have addressed the relationship between the mass surface densities of the cold neutral (molecular and atomic) gas in galaxies and their star-formation rate surface densities. This includes above all the seminal work of \citet{schmidt59} and \citet{kennicutt89}, who established a single, tight ($\sim 0.3$~dex), empirical star-formation law which governs the efficiency with which stars form from cold neutral gas in galaxies over more than five orders of magnitude. While there is a broad consensus that this relationship heralds a close physical link between gas and star formation, the detailed physical processes underlying this relationship are still a matter of active debate. 

Moreover, alternative relationships have been proposed for the star-formation rate surface density, in particular with the ratio of gas surface density to dynamical (disk rotational) time \citep[][]{kennicutt98}. More recently, \citet{shi11} and \citet{shi18} pointed out  that a relationship which not only accounts for the surface density of gas, but also of stellar mass, remains valid also for star-formation at very low metallicities and gas mass surface densities, where the classical star-formation law breaks down. 

Finding significant, systematic offsets in AGN hosts from the star-formation law of purely star-forming galaxies would be one signature indicating that AGN feedback can indeed affect star formation in their host galaxies. This would be more constraining than the presence of outflows, because it may indicate a direct physical link between the injection of AGN jet energy into the gas and a suppression of star formation and galaxy growth. Several studies of other classes of AGN have not identified a deficit in star formation in spite of the presence of fast outflows \citep[e.g.,][]{genzel14,harrison16}. 

\citet{nesvadba10} and \citet{ogle10} were the first to show that the star formation intensity at a given gas mass surface density (star-formation efficiency) may be unusually low in the subset of early-type radio galaxies which are rich in relatively dense molecular gas, including gas that is heated by shocks. They found a characteristic offset from the usual Kennicutt-Schmidt relationship by about a factor~10 relative to other galaxy populations. This result was later on confirmed by Herschel far-infrared photometry of dust heated by star formation \citep[][]{lanz16}. 

These authors initially attributed the offset towards lower star
formation efficiencies to the gas being perturbed by the radio jet
\citep[][]{nesvadba11}, a process which might also create and maintain
the large reservoirs of warm molecular hydrogen found in these
galaxies, which appear to be heated by shocks
\citep[][]{ogle07,nesvadba10,ogle10}. 

However, this interpretation was
subsequently challenged by \citet{martig13} and \citet{saintonge12},
who proposed that the reason for low star-formation efficiencies in
early-type galaxies was inherent to the spheroidal morphology of their
stellar component, which would stabilize the gas against becoming
gravitationally bound and forming stars. In this case, the presence of
radio jets and turbulent gas would just be a coincidence, with no
direct physical link to the star formation. 

Comparing the star-formation efficiency in J2345$-$0449 with that
of radio galaxies of \citet{ogle10} is very
interesting, because the molecular ring in J2345$-$0449 is associated
with the disk, not the bulge of the galaxy. Hence the high stellar
mass surface density or the morphological quenching invoked by
\citet{martig09,martig13} cannot regulate the star formation in this
gas. Demonstrating that the gas and star
formation density in J2345$-$0449 fall near the \citet{ogle10} radio galaxies
would therefore demonstrate that a radio source alone is able to
produce such an offset in star-formation efficiency, even in the absence
of the high stellar mass surface densities and spheroidal morphologies
typically found in early-type galaxies. A potential impact of angular momentum and stellar mass surface density can be further investigated by comparing J2345$-$0449 with the alternative star-formation laws. Finding significant offsets in all these relationships would be an important
piece of evidence to demonstrate that AGN feedback is a viable
mechanism to lower star-formation rates in massive, gas-rich galaxies. 

\subsubsection{Standard Kennicutt-Schmidt relation}
We will now use the star-formation rate
surface densities from Section~\ref{ssec:starformation}, and the gas-mass surface densities from Section~\ref{ssec:sigmagas}, to obtain average surface densities in both
quantities. Currently available data can only provide us with average
values, because GALEX has a relatively low spatial resolution of only
5\arcsec. However, as already discussed in detail in Section~\ref{ssec:starformation}, both the UV and 22~$\mu$m continuum morphologies of J2345$-$0449 are spatially resolved and show that star formation is extended over the entire stellar disk, with very similar, low, integrated star-formation rates of SFR=$1-1.25\ M_{\odot}$ yr$^{-1}$. This implies that GALEX does not systematically miss a strong star-forming component deeply embedded in dust.  

\begin{figure}
  \includegraphics[width=0.49\textwidth]{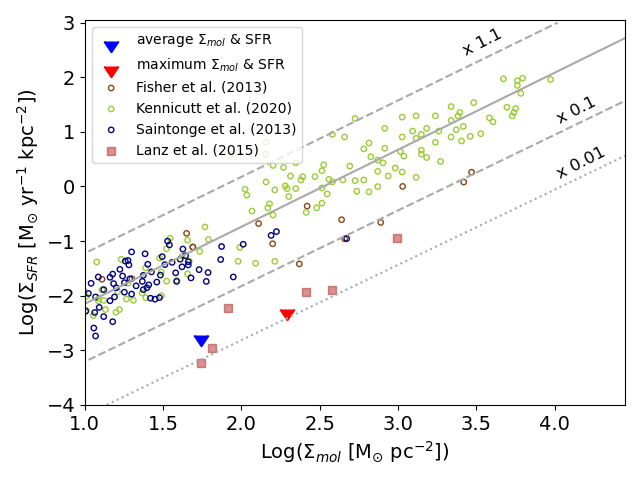}
  \caption{\label{fig:skplot}
  Star-formation rate
    surface density as a function of molecular gas-mass surface
    density in J2345$-$0449. The blue and red upside-down triangles indicate the average and maximal gas-mass surface densities measured from CO(1--0), respectively. Values of star formation indicate the maximal and minimal local rates given in Section~\ref{ssec:starformation}.}
\end{figure}

Fig.~\ref{fig:skplot} shows where J2345$-$0449 falls in the
Kennicutt-Schmidt diagram with respect to low-redshift spiral and
starburst galaxies, and the radio galaxies of \citet{ogle10} and
\citet{lanz16}, for which spatially resolved measurements of CO line
emission are available. For J2345$-$0449 we plot the average and
maximal gas-mass surface densities derived from our ALMA
interferometry (Section~\ref{ssec:sigmagas}), and the minimum and maximum of local star-formation densities as given in Section~\ref{ssec:starformation}, $1.3$ and $5.6\times 10^{-3}$ M$_{\odot}$ yr$^{-1}$ kpc$^{-2}$, respectively. These two
data points are therefore not unique values derived for a specific pair of local surface density of
star formation and molecular gas, but they are chosen to illustrate a representative range of surface densities of molecular gas mass and star formation in J2345$-$0449, respectively.  

For both, the galaxy falls well below the standard Kennicutt-Schmidt relationship and into the regime that is set by the radio galaxies of \citet{ogle10}, \citet{nesvadba10}, and \citet{lanz16}. Taken at face value, local star formation rates are factors 50 to 75 lower than what is expected from the standard Kennicutt-Schmidt relation at the given gas mass surface densities. Throughout the paper, we have already argued that this cannot be the result of the same structural properties as for bulges, because J2345$-$0449 has no prominent bulge, no unusual stellar mass surface densities compared to other galaxies, and no dense environment. The unusually bright X-ray halo compared to other spiral galaxies in the same mass range, and presence of $2\times 10^{10}$ M$_{\odot}$ of cold molecular gas also show that shock heating of infalling gas at the virial radius alone has not been able to inhibit the accumulation of significant amounts of gas in this galaxy. 

\subsubsection{Alternative relationships}
\begin{figure}
  \includegraphics[width=0.49\textwidth]{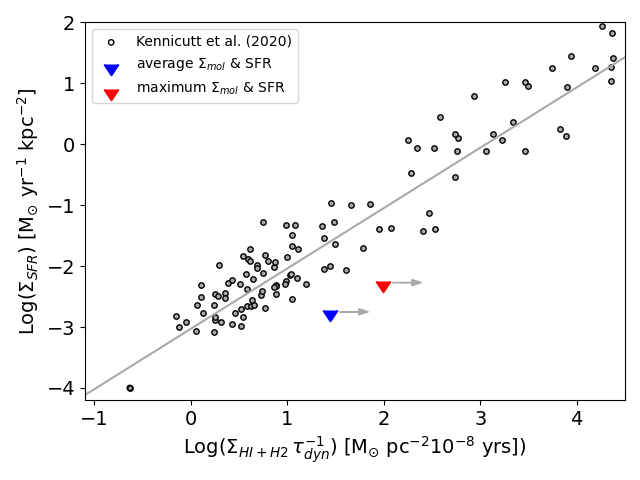}
  \caption{\label{fig:skplot_time}
  Star-formation rate
    surface density as a function of the ratio of molecular gas-mass surface
    density and dynamical (rotational) time in J2345$-$0449, for the average and maximal gas-mass surface densities measured from CO(1--0), and assuming that the
    star-formation rate of $SFR=1.25$ M$_{\odot}$ yr$^{-1}$ is uniformly distributed across the disk as indicated by the GALEX and WISE morphologies  (Section~\ref{ssec:starformation}). Both measurements of gas-mass surface density in J2345-0449 are lower limits, because we have no HI measurement. Nonetheless, the molecular gas mass surface densities alone already show a pronounced offset towards higher gas-mass surface densities than expected for normal star-forming spirals.}
\end{figure}

\begin{figure}
  \includegraphics[width=0.49\textwidth]{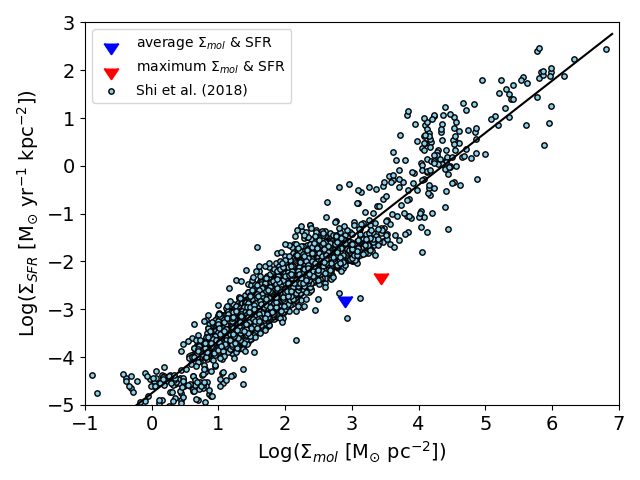}
  \caption{\label{fig:skplot_mass}
  The extended star-formation law for J2345-0449 and the sample of 
  \citet{shi18} shown for comparison. Although J2345-0449 falls closer to the overall galaxy population than for the classical Schmidt-Law and the Silk-Elmegreen law, a marked offset of $\ge1$~dex towards lower star-formation rate surface densities persists (see text for details).} 
\end{figure}

In Figs.~\ref{fig:skplot_time} and Figs.~\ref{fig:skplot_mass}, we investigate whether J2345$-$0449 is more typical in alternative star-formation laws, like the Silk-Elmegreen law between star-formation rate surface density and gas-mass surface density divided by the dynamical time \citep[][]{kennicutt98b}, and the extended Schmidt Law \citep[][]{shi11,shi18}, which takes into account the effect of stellar mass surface density by relating star-formation rate densities with the product of the gas mass surface density, $\Sigma_{gas}$ and square root of the stellar mass surface density $\Sigma_{stellar}^{1/2}$, i.e. 200~M$_{\odot}$~pc$^{-1}$ in J2345$-$0449 (Section~\ref{ssec:sigmastars}). This quantity scales with the hydrostatic mid-plane pressure of the galaxy. For neither relationship does J2345$-$0449 fall onto the ordinary relationship between gas and star formation rate density. The offset from the Silk-Elmegreen relationship is smaller than that in the Kennicutt-Schmidt relationship, however, our result should be considered a lower bound. The comparison sample uses the masses of cold molecular and atomic gas combined. In J2345$-$0449, the molecular gas mass alone is already sufficient to produce an offset of 0.65~dex from the sequence of ordinary galaxies. An offset that can only increase if cold atomic gas was included. 

The extended Schmidt-law overpredicts the star formation rate surface density in J2345$-$0449 as well, by 1.0~dex for average, and by 1.3~dex (a factor 21) for the highest mass surface densities. We conclude that J2345$-$0449 falls off all three star-formation laws, as would be expected if star formation was regulated by a mechanisms which is not present in the overall galaxy population. 

\subsection{Rotational support in the ring with and without radio jet}
An additional hypothesis to explain the low star-formation efficiency in J2345$-$0449, which is specific to massive spiral galaxies with high angular momentum compared to bulges, is that the high rotational velocity of the ring could make the gas rotationally stable at the observed gas-mass surface densities \citep[e.g.,][]{ogle19}. It is therefore very instructive to contrast the results of such an analysis for gas without and with line broadening caused by the radio source. 

We use the \citet[][]{toomre64} criteria to estimate the minimal gas-mass surface density which would allow the gas in the ring in J2345$-$0449 to become self-gravitating, fragment, and form stars. Following, e.g., \citet{toomre64} or \citet{martin01}, the critical mass surface density, $\Sigma_{gas,crit}$ for gas to become Toomre-unstable is $\Sigma_{gas,crit}=\kappa\ \sigma / \pi\ G$, where $\sigma$ is the Gaussian line width, $\sigma=FWHM/2.355$, $G$ the gravitational constant, and $\kappa$ the epicyclic frequency. We follow \citet{genzel14} in setting $\kappa=a v_c/R$. $v_c$ is the circular velocity at a given galactocentric radius, $R$ is that radius, and $a$ is a factor that takes into account the geometric properties in the disk. We adopt $a=1.4$, appropriate for the flat part of the rotation curve. 

Adopting an appropriate value for the gas velocity dispersion in absence of the radio jet is somewhat delicate, as line widths are very broad in most of the ring. The lowest values found in the eastern and western ring, about orthogonal to the jet axis, are around $\sigma=10$ km s$^{-1}$ (Fig.~\ref{fig:almamaps}), similar to those found in other massive spiral galaxies without prominent radio sources \citep[e.g.,][]{tamburro09}. We will therefore adopt $\sigma=10$ km s$^{-1}$ as typical value for the gas velocity dispersion in absence of the radio source.

With these definitions, setting $R_{inner}=4.2$~kpc and $R_{outer}=11.8$~kpc for the inner and outer edge of the molecular disk, respectively, and using the deprojected rotation velocities $v_{c,inner}=366$~km s$^{-1}$ and $v_{c,outer}=378$ km s$^{-1}$ measured at these radii, respectively, we find critical gas mass surface densities of 90.5~M$_{\odot}$ pc$^{-2}$ in the inner, and 34.0 M$_{\odot}$ pc$^{-2}$ in the outer ring. Many authors, including in particular \citet{martin01}, have found that normally star-forming galaxies have gas-mass surface densities which are between about 0.6 and 1.5$\times$ the critical value. The same is the case for most of the gas in the molecular ring of J2345$-$0449 when adopting typical gas velocity dispersions in galaxies with $M_{stellar}=few \times 10^{11}$~M$_{\odot}$, $10$ km s$^{-1}$, and the average, median, and maximal gas mass surface densities of 55, 83, and 196 $M_{\odot} pc^{-2}$, respectively (Section~\ref{ssec:sigmagas}).

It therefore seems that the rapid ring rotation alone cannot explain why star-formation rates in J2345$-$0449 are so much lower than in other spiral galaxies. However, this changes, of course, when considering the measured line widths instead of those measured in absence of powerful radio jets. For example, at $\sigma=30$ km s$^{-1}$, the threshold value we adopted to uniquely identify gas affected by the radio jet in Section~\ref{ssec:outflows}, or at $\sigma=75$~km s$^{-1}$, corresponding to the highest line widths measured in the ring, critical gas mass surface densities in the outer ring are $\Sigma_{gas,\ crit}=\ 102\ M_{\odot}$~yr$^{-1}$ and $\Sigma_{gas,\ crit}=\ 255\ M_{\odot}$~yr$^{-1}$, respectively. At the inner edge of the ring, they would be ever higher. This illustrates that the line broadening from the radio jet may well make gas in an otherwise marginally rotationally supported gas disk unable to fragment into individual, self-gravitating, and star-forming gas clouds. A more detailed analysis would require higher-resolution constraints on the star formation in J2345$-$0449, and will become possible once our scheduled MUSE observations have been taken.     

\section{Discussion}
\label{sec:discussion}
\subsection{A case for jet-ISM interactions in the molecular ring}\label{sec:simcompare}
\begin{figure*}
\centering
  \includegraphics[width=0.32\textwidth]{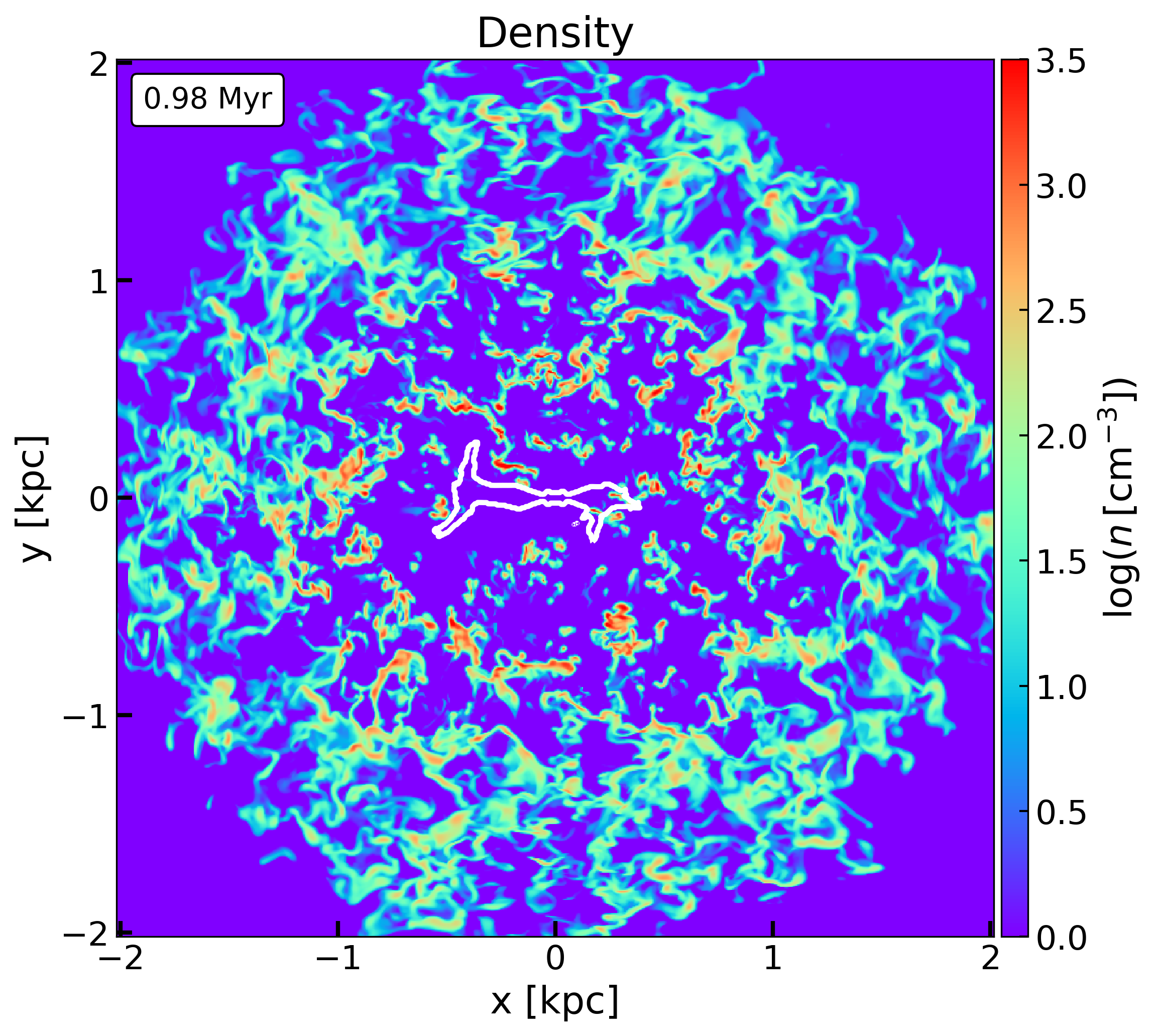}
  \includegraphics[width=0.32\textwidth]{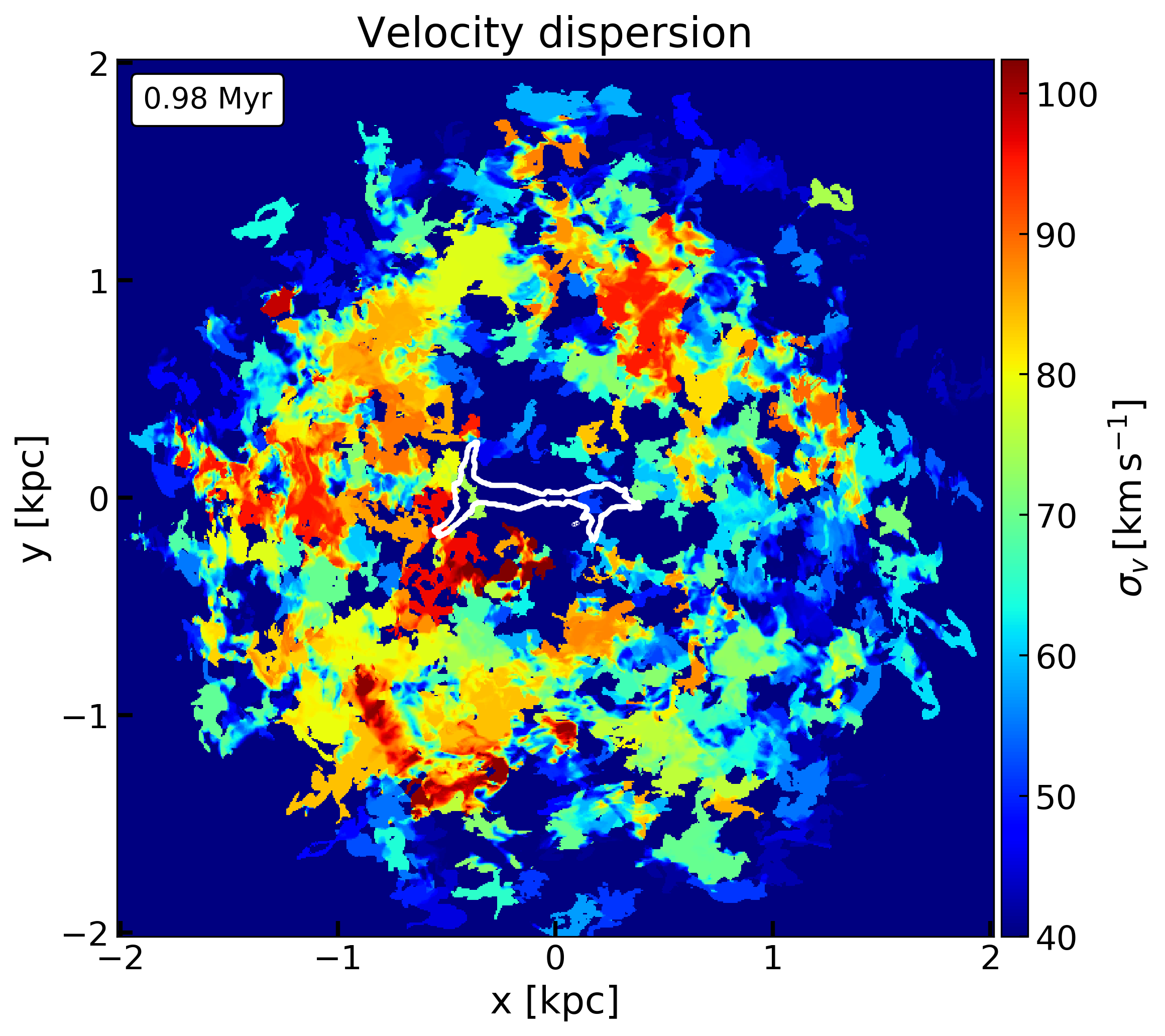}
  \includegraphics[width=0.32\textwidth]{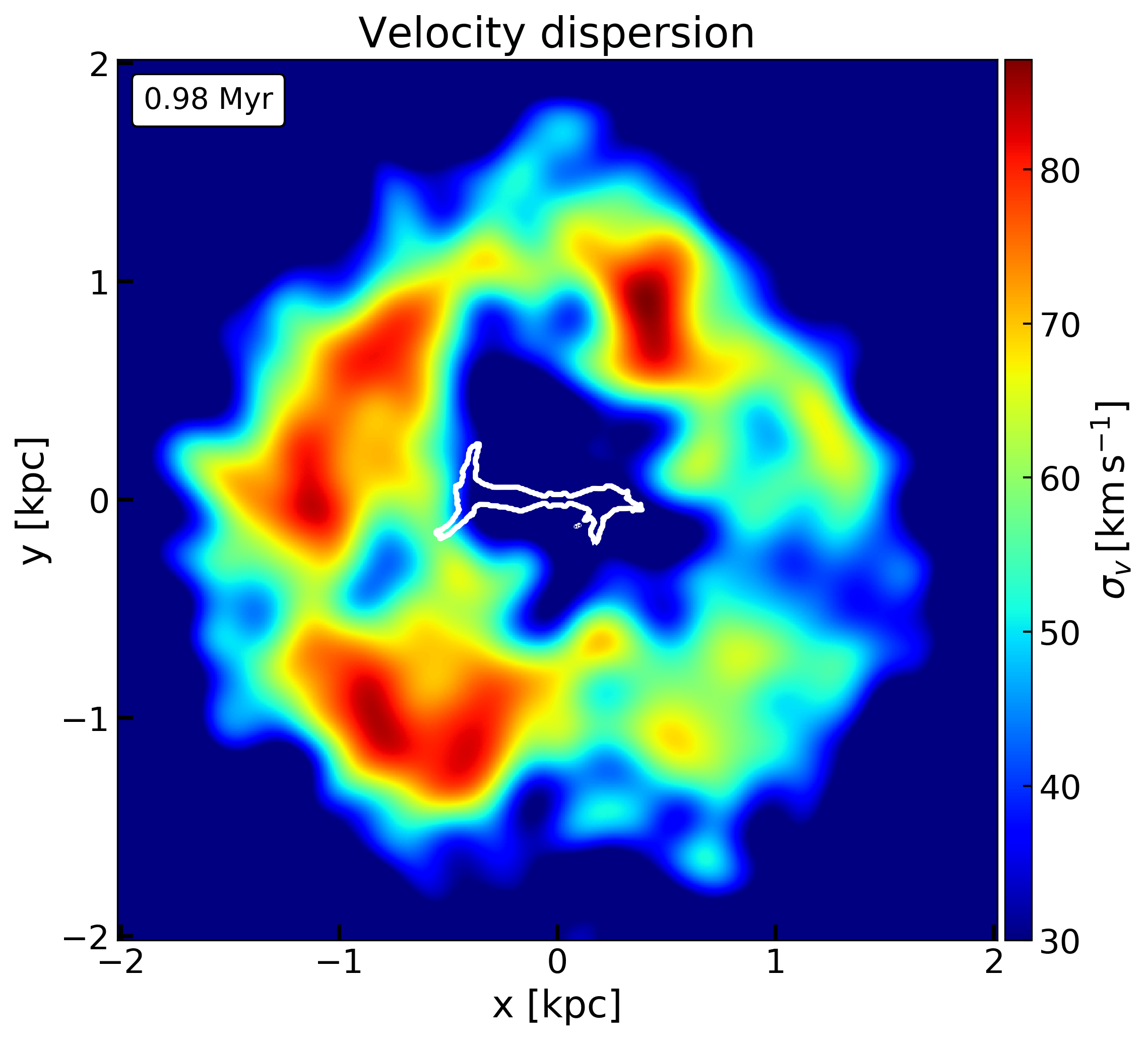}
  \caption{\label{fig:simDplots} {\it Left:} A density slice through the disk mid-plane (X-Y plane at Z=0) for simulation D of \citet{mukherjee18}. The white lines denote the contour of value 0.5 of the jet tracer, projected onto the X-Y plane. {\it Middle:} Mean velocity dispersion of dense gas in the X-Y plane. {\it Right:} Mean velocity dispersion smoothed with a Gaussian kernel of width 100 pc. See Sec.~\ref{sec:simcompare} for further details.}
\end{figure*}
It is instructive to compare these observational results for J2345$-$0449 with existing simulations of the interactions between
relativistic jets and the interstellar medium.
\citet{mukherjee18} performed a set of relativistic hydrodynamic simulations of radio galaxies with a range of jet power, gas conditions, and jet orientations relative to the gas disk to explore the propagation and interaction of the radio jet with an inhomogeneous, turbulent, multi-phase interstellar medium on kpc scales. Although these simulations were not conducted to model a specific radio galaxy, they demonstrate how jets heat, disperse, and drive out warm and cold gas. Hence, they can help us make 
a qualitative comparison with the data presented in this paper. In the following, we  use their simulation D, which describes the signature of a  radio jet with power $P_{jet}=10^{45}$ erg s$^{-1}$ impinging at an inclination angle of 45$^\circ$ on a 4 kpc diameter gas disk with mass $M_{gas}=5.7 \times 10^9\ M_{\odot}$. For the full list of simulation parameters, see \citet[][Tables~1 and 2]{mukherjee18}. Since the jet is inclined by 45$^\circ$ to the disc, this simulation most closely resembles the geometry of the jet-disc system in J2345$-$0449.

In Fig.~\ref{fig:simDplots} we illustrate the impact of jet-ISM interactions on the kinematics of the dense gas in the simulation. In the left panel, we present a top-view of the density map of simulation D at the mid-plane of the disk (Z=0 plane).  The location of the jet, projected onto the X-Y plane, is shown in white.
The middle panel shows the velocity dispersion of dense clumps with $n > 100\mbox{ cm}^{-3}$ 
projected onto the X-Y plane by performing a mass-weighted average along the Z axis. Further numerical details will be discussed in a forthcoming publication (Mandal et al. in prep). We also show a smoothed image of the velocity dispersion by convolving the original map at the resolution of the simulation ($\delta x \sim 6$ pc), with a Gaussian kernel of width $\sigma = 100$ pc. The convolved image, representing a velocity dispersion  map viewed at a poorer resolution, still shows inhomogeneous structures, with enhanced dispersion at the inner edges of the central cavity. The morphology of the dispersion maps in Fig.~\ref{fig:simDplots} obtained from the simulations, bear striking resemblance to the observed distribution of the width of CO(1-0) lines presented in Fig.~\ref{fig:almamaps}, although on differing spatial scales.

It is evident from Fig.~\ref{fig:simDplots} that one of the strongest points of interaction between the jet and the disk is near the jet head (to the left of the image), where sustained head-on collisions between the jet streams and clouds occur. This raises the velocity dispersion of the dense gas in the immediate vicinity to $\sigma \gtrsim 70\ \mbox{km\ s}^{-1}$, as a substantial fraction of jet kinetic energy
is transformed into cloud motions.
Strong thermal pressure gradients and ram pressure gradients are carried away from the head by secondary jet streams and act in all directions, dispersing and heating gas in a large region around the jet head. The velocity dispersion in the entire gas disk increases as a result of the jet-ISM interactions to values larger than $\sigma \gtrsim 50\ \mbox{km s}^{-1}$, higher than the starting dispersion of the unperturbed disk at the beginning of the simulation.

The qualitative similarity between the simulated gas kinematics discussed above and the observed results in Fig.~\ref{fig:almamaps}, prompts us to infer that the enhanced turbulent dispersion observed in J2345$-$0449 results from active interaction of the jet-driven flows with the gas disk. While the accelerated and dispersed clouds, especially near the head, are strongly compressed and ablated, very little gas reaches escape velocity, as also observed in J2345$-$0449. Instead, the gas participates in a fountain-like outflow, that could further contribute to raising the gas turbulence to the high values observed in J2345$-$0449 when falling back onto the disk. 

In spite of the mass loss through the fountain, after $\sim 1$ Myr, the central 0.5 kpc region of the disk is still only partially cleared of gas. Judging from the dynamics of the dense gas being pushed away by the jet, which remains confined for the entire $\sim 2$ Myr of the simulation, we expect that by 5 -- 10 Myr a sizeable cylindrical hole devoid of dense gas will be created within the inner 0.5 kpc of the gas disk, as also seen, e.g., in the simulation by \citet{gaibler11}. Apart from the excavated central region, the structure and rotation of the gas disk outside the central cavity remain largely intact because the jet-ISM interactions do not remove a sufficient amount of angular momentum to disrupt the entire disk. This suggests that the gas morphology in J2345$-$0449 can be a result of radio jet feedback, even if the galaxy hosted a gas disk extending all the way to the central regions before the onset of the radio jet activity. The absence of large amounts of dense gas is also consistent with the radio source being fueled by accretion of hot gas onto the central supermassive black hole \citep[e.g.,][]{hardcastle07}. 

Strong jet-ISM interactions also lead to asymmetric radio morphology \citep{gaibler11} as appears to be the case for J2345$-$0449. 
For the duration of the simulation, the main jet stream is deflected toward the cylindrical axis of symmetry of the gas disk and escapes nearly perpendicular to the disk. Thus, while the jet is inclined by 45$^\circ$ at the base and within the galactic disk, at larger distances it propagates perpendicular to the disk. In the case of J2345$-$0449 the deflection is seen in projection and whether the jet has broken out of the disk and is freely propagating into the galactic halo, or whether it is still being deflected is not immediately discernible without a thorough analysis of the radio data and, possibly, not without new radio data at better sensitivity and higher angular resolution. 

A simulation tailored to J2345$-$0449 would be required for a more quantitative comparison. However, even a qualitative comparison as presented here already allows us to draw general conclusions about the nature of the interaction between jet and gas in J2345$-$0449. The impact that such interactions can have on the star formation in the host galaxy have so far only started to be investigated with simulations \citep[][]{gaibler12, bieri16, mukherjee18}, sometimes with somewhat crude assumptions. We are currently performing a more detailed study of star formation in the presence of radio jet feedback, which shows that such such an interaction can indeed lower the star-formation efficiency of a massive gas disk, as seen in J2345$-$0449 (Section~\ref{sec:sflaw}). Such an analysis is beyond the scope of this paper and will be presented in a separate publication (Mandal et al., in prep.).

\subsection{From the current feedback episode to the global suppression of gas cooling and galaxy growth}
\label{ssec:xray}

Extrapolating from in-situ observations of radio jet feedback to the global implications of such episodes for the growth history of massive galaxies is one of the most important aspects of AGN feedback studies, but observational constraints that allow us to directly quantify both in the same galaxy are usually very difficult to come by. Past X-ray observations of J2345$-$0449 by \citet{walker15} and \citet{mirakhor20} revealing a bright X-ray halo surrounding the disk, and the highly unusual presence of a radio jet in this late-type spiral give us the rare opportunity to discuss complementary observational constraints on the on-going feedback episode seen in molecular gas, and on the global suppression of star formation derived from the baryon budget in stars and hot halo gas.

As \citet{mirakhor20} recently pointed out, the high baryon fraction of 60\% of the cosmic average in the X-ray halo of J2345$-$0449 \citep[for a cosmic baryon fraction of 18.8\%,][]{planck18} suggests that the underlying, massive dark-matter halo of $1.07\times 10^{13} M_{\odot}$ \citep[][]{mirakhor20} must have retained a large fraction of its infalling gas. 
This is very interesting, because it implies that the baryons in J2345$-$0449 must also carry the signatures of the integrated AGN feedback history, if such feedback is to play a major role for regulating the stellar growth of massive galaxies, as is now widely believed. We will now compare the radio and X-ray properties and stellar mass fraction of J2345$-$0449 with those of other massive spiral galaxies to constrain the potential global impact of radio jet feedback on its integrated star-formation history. 

\subsubsection{J2345$-$0449 in the context of massive spiral galaxies}

As a class, massive spiral galaxies akin to J2345$-$0449 have only recently received
increased attention in the literature. While it is not uncommon that these
galaxies have low specific star-formation rates compared to less
massive spiral galaxies, \citet{posti19} recently showed that they are also unusually rich in stellar mass for their dark-matter halo mass, compared to expectations from cosmological abundance matching. They present detailed fits of the HI rotation
curves and mass-to-light ratios to dark-matter halo profiles of 175
spiral galaxies with stellar masses of $M_{stellar}=$~few$\times10^{9-11}
M_{\odot}$ from the SPARCS sample \citep[][]{lelli15}. The baryon
fraction in their sample increases monotonically with increasing
stellar and dark-matter halo mass, up to their most massive galaxies with $M_{stellar}=\ {\rm few}\times\ 10^{11}$~ M$_{\odot}$, akin to
J2345$-$0449. This is in contrast to cosmological abundance matching studies, which suggest a maximal efficiency of baryon cooling and star formation in galaxies at stellar masses 
$M_{stellar}\sim 10^{10.5}$ M$_{\odot}$ \citep[e.g.,][]{moster13,behroozi13}. \citet{posti19} suggest that the difference between these results could be explained if AGN feedback was the primary reason why the stellar mass growth of these galaxies, mainly early-types, became increasingly inefficient: If black-hole growth and bulge growth are related, they argue, then baryon cooling in massive spirals, which would generally not experience strong AGN feedback, would become increasingly efficient compared to early-type galaxies. In broad support of this statement, \citet{li19} did not find evidence for significant offsets from the Kennicutt-Schmidt relationship in massive spiral galaxies with millimeter CO emission-line observations.

\begin{figure}
  \centering
  \includegraphics[width=0.49\textwidth]{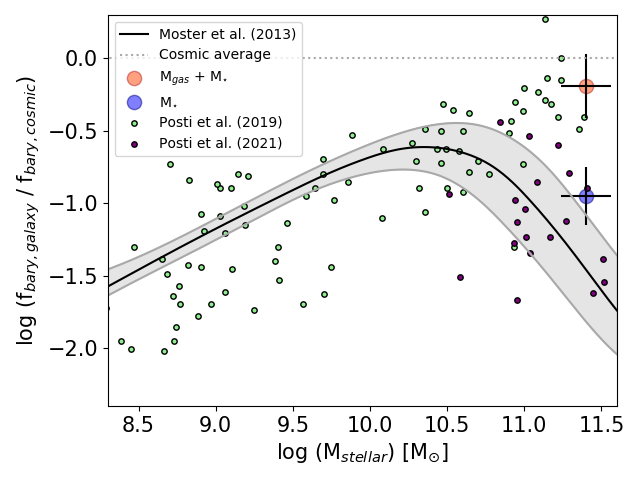}
  \caption{\label{fig:fbary} 
    Comparison of the observed baryon fraction
    in galaxies, scaled by the cosmic baryon fraction. 
     For J2345$-$0449 we show the ratio of stellar mass to dark-matter halo mass (blue stars). The red star shows the ratio of baryonic mass (the sum of stellar mass, and mass of hot halo and molecular gas) to dark-matter halo mass. Values for other galaxies are taken from \citet{posti19} and \citet{posti21}, and include only stellar mass, $f_{bary,galaxy}=M_{stellar}/M_{DM}$. Even without including halo gas, they fall very near the cosmological limit of $f_{bary,cosmic}=\Omega_{baryon}/\Omega_{DM}=0.188$, indicating that they have already transformed most of their infalling gas into stars. This is not the case for J2345$-$0449, which has about $2/3$ of its baryons in the hot halo.} 
\end{figure}

J2345$-$0449 has a high baryon fraction of 60\%, comparable to the galaxies of \citet{posti19}. However, with a stellar mass of $M_{stellar}=3.6\times 10^{11}$ M$_{\odot}$ (Section~\ref{ssec:stellarmass}) and a mass of hot halo gas, $8.25\pm^{1.62}_{-1.77}\times 10^{11} M_{\odot}$, only 30\% of the baryonic mass is in stars and cold gas in this case, the remaining 70\% is in the hot halo, whereas the high baryon fractions in the massive spirals of the \citet{posti19} sample are derived from the stellar mass alone. 

In Fig.~\ref{fig:fbary}, we reproduce the upper panel of Fig.~2 of \citet{posti19} to illustrate where J2345$-$0449 falls relative to their sample and to expectations from abundance matching by \citet{moster13}. If counting only the stellar mass in J2345$-$0449, then its baryon fraction falls a factor of~$\ge3$ short of equally massive galaxies in the \citet{posti19} sample, and into the same realm as early-type galaxies, whereas it has a rather typical baryon content akin to other massive spiral galaxies, when including also the hot X-ray gas. The powerful radio source is the most striking difference between J2345$-$0449 and the \citet{posti19} sample, and is affecting the cold interstellar gas and star formation, as we have shown before, as well as the hot halo gas, as previously demonstrated by \citet{walker15}. It is therefore reasonable to consider the difference in stellar mass fraction between J2345$-$0449 and the \citet{posti19} sample the long-term consequence of radio jet feedback. J2345$-$0449 seems to have accreted about as much baryonic mass as other spiral galaxies in the same mass range, however, about two thirds of these baryons do not seem to have formed stars, but are still (or again) in the halo. 

\subsubsection{Is radio jet feedback at the origin of the low stellar mass fraction in J2345$-$0449?}
Would the current kinetic jet power observed in J2345$-$0449 be sufficient to explain the deficit in stellar mass compared to the massive spiral galaxies of \citet{posti19}? 
For simplicity we will assume that the current feedback episode in J2345$-$0449 is typical, and that without a supermassive black hole capable of launching radio jets, the galaxy would have transformed all of the $8.25\times\ 10^{11}$ M$_{\odot}$ in hot halo gas into stars. \citet{sabater19} suggest that galaxies with at least $4\times 10^{11}$ M$_{\odot}$ in stellar mass and radio sources at least as powerful as in J2345$-$0449, $E_{kin}=2\times 10^{44}$~erg~s$^{-1}$, have duty cycles of about 0.1, i.e., they host radio sources at least as powerful as that in J2345$-$0449 for about 10\% of the time. This would suggest that the jet of J2345$-$0449 has released about $6\times 10^{60}$ erg into the interstellar and circumgalactic gas in the last 10~Gyrs. Following \citet{walker15}, lifting gas out of the galaxy and to radii beyond the bright inner X-ray halo (80~kpc) would require injecting about $5\times 10^{48}$ erg s$^{-1}$~M$_{\odot}^{-1}$ of kinetic jet energy into the halo, suggesting that the jet could have heated and even unbound up to about $1.2\times 10^{12}$~M$_{\odot}$ in the past 10~Gyrs. This does not take into account, e.g., gas cooling or inefficiencies in the energy transfer from the jet into the gas, however, it does plausibly suggest that energy deposition through the radio jet into the gas  alone may explain the relative deficit in stellar mass in J2345$-$0449 compared to the \citet{posti19} sample. Finding multiple pairs of bright radio jets in J2345$-$0449 \citep[][]{bagchi14} gives direct evidence of a long-term jet activity in this source (Section~\ref{ssec:kinenergy}). 

The low star-formation efficiencies found in Section~\ref{sec:sflaw} suggest that star formation can be inhibited during AGN episodes, even if not all gas is immediately outflowing in a rapid, instantaneous blowout event, as has initially been suggested by early cosmological models of AGN feedback. The molecular gas mass is comparable to the hot X-ray gas within the inner $\sim 80$~kpc of the halo \citep[][]{walker15}, and corresponds to $\sim 2.5$\% of the total hot gas content in the halo as estimated by \citet{mirakhor20}. Much of the disk gas may have accumulated through partial cooling of the hot gas out of the halo, in analogy with X-ray bright galaxy clusters  \citep[][]{revaz08,voit15,gaspari18}. \citet{walker15} estimated that a large fraction of the jet kinetic energy is likely necessary to inflate X-ray cavities in the hot halo gas; in comparison, the energy injection rates of only a few percent of the jet energy, that are required to keep the cold gas in the host galaxy from forming stars (Section~\ref{sec:sflaw}), are rather low. This suggests that inhibiting star formation in the molecular gas of the host galaxies of radio-loud AGN may be possible even with a comparably modest energy input compared to the overall feedback cycle in radio-loud AGN host galaxies. Nonetheless, it has major consequences for the overall stellar growth history of these galaxies, independent of the morphological or structural properties of these galaxies themselves. 

\section{Summary}
\label{sec:summary}
We have presented an analysis of new ALMA CO(1--0) interferometry of the massive spiral galaxy 2MASS~J23454368$-$0449256 (J2345$-$0449) at $z=0.0755$, which is exceptional for its multiple pairs of bright radio lobes reaching out to a maximum of 1.6~Mpc, and its bright X-ray halo \citep[][]{bagchi14,walker15}. 
We detect a 24~kpc wide ring associated with the inner stellar disk of the galaxy, which contains $2.0~\times10^{10}$~M$_{\odot}$ of cold molecular gas. Line emission in about 3/4 of the ring surface has line widths that are FWHM$\ge 70$~km~s$^{-1}$, well above typical widths of CO line emission in other, equally massive, spiral galaxies without radio AGN. These regions are also intercepted by the jets, and we therefore interpret the broadening as due to feedback from the radio source. We do not detect any outflow motion at velocities exceeding the circular velocity of the ring, $v_c=369~\pm\ 25$~km~s$^{-1}$. 

Molecular gas-mass surface densities are between 55 and $196\ M_{\odot}$ pc$^{-2}$. We estimate stellar mass surface densities for the disk and low-mass pseudo-bulge, finding that they are typical for bulges and disks in other spiral galaxies, including spiral galaxies at lower mass. Star-formation rate surface densities derived from spatially resolved GALEX far-UV imaging are $1.3-5.4\times 10^{-3}$ M$_{\odot}$ yr$^{-1}$~kpc$^{-2}$, $50-75\times$ lower than in other spiral galaxies without prominent radio source and similar gas-mass surface densities, assuming that a common CO-to-H$_2$ conversion factor applies to all galaxies. This places J2345$-$0449 well below the ordinary Kennicutt-Schmidt law and into the same realm already populated by powerful early-type radio galaxies, which are rich in molecular gas, including warm molecular gas presumably heated by shocks induced by their radio jets. Significant offsets towards lower star-formation rate densities are also found for the Silk-Elmegreen relationship, and the extended Schmidt-law, which takes into account stellar mass surface density. 

We use the Toomre stability criterion to show that the molecular gas in J2345$-$0449 would be near the critical mass surface densities where rotational support would make the gas marginally stable to forming self-gravitating clouds and stars, if line widths were as low as in the small regions of the ring not affected by the radio source. These regions of quiescent gas in J2345$-$0449 have line widths that are comparable to those in other, equally massive spiral galaxies, which fall onto the ordinary star-formation laws.  

However, the increase in gas velocity dispersion caused by the radio jet in J2345$-$0449, and corresponding to about 1-2\% of the kinetic energy carried by the radio jets, enhances the critical densities to values that are significantly higher than those found in most parts of the ring. We find that alternative processes, like heating by the bolometric AGN radiation, star-formation, but also structural properties of the host galaxy as often suggested for the by far more typical early-type host galaxies of powerful radio sources, can be ruled out as causes of the low star formation rates in this galaxy, which has a low-mass pseudo-bulge, no bar, and no rich environment. This suggests that interactions with the radio jet, as observed here, are indeed the cause of the low star formation efficiency in J2345$-$0449, and not these other processes, which are commonly co-existent with powerful radio jets in more common bulge-dominated and early-type host galaxies of radio jets. 

J2345$-$0449 is an outstanding object for detailed studies of the consequences of radio AGN feedback for star formation in galaxies generally. We also argue that this impact may be long-term, as J2345$-$0449 seems to have retained large fractions of its baryons in its massive dark-matter halo, like other massive spiral galaxies recently studied, which have no radio jets. However, most of these baryons seem to be in hot halo gas, and not in stellar mass in the host galaxy. This is a central prediction of the AGN feedback theorem in early-type galaxies. Finding the same to be the case in J2345$-$0449, a rare spiral galaxy with powerful radio source, but not in other, equally massive spiral galaxies without jets, adds important further credence to this mechanism. 

\bibliographystyle{aa}
\bibliography{j2345}

\acknowledgements
The authors wish to thank L. Posti and P. Dabhade for interesting discussions about stellar to halo mass fractions and previous observations of this source, and also thank the staff at the IRAM 30-m telescope for their support during our remote observing program. R.M.J. Janssen is supported by an appointment to the NASA Postdoctoral Program at the NASA Jet Propulsion Laboratory, administered by Universities Space Research Association under contract with NASA. This work is mainly based on the following ALMA data: ADS/JAO.ALMA\# 2019.1.01492.S. ALMA is a partnership of ESO (representing its member states), NSF (USA),and NINS (Japan), together with NRC (Canada), NSC, and ASIAA (Taiwan), and KASI (Republic of Korea), in cooperation with the Republic of Chile. The Joint ALMA Observatory is operated by ESO, AUI/NRAO, and NAOJ. 
\end{document}